\documentclass{aa}  
\usepackage{graphicx}
\usepackage[varg]{txfonts}
\usepackage{natbib}
\usepackage{hyperref}

\newcommand\un[1]{{\,\rm #1}}
\newcommand\E[1]{\times10^{#1}}
\newcommand\rs[1]{_\mathrm{#1}}
\newcommand\pd[2]{\frac{\partial{#1}}{\partial{#2}}}

\usepackage{ulem}
\usepackage[dvipsnames]{xcolor}

\begin{document}

\title{On the acceleration of cosmic rays at the post-adiabatic shocks of supernova remnants}
\titlerunning{CR acceleration at post-adiabatic shock}
\author{O. Petruk\inst{1,2}
    \and
    R. Bandiera\inst{3}
	\and
	T. Kuzyo\inst{2}
	\and
    R. Brose\inst{4}
    \and
    A. Ingallinera\inst{5}
}

\institute{INAF - Osservatorio Astronomico di Palermo, Piazza del Parlamento 1, 90134 Palermo, Italy\\
    \email{oleh.petruk@inaf.it}	
	\and
	Institute for Applied Problems in Mechanics and Mathematics, National Academy of Sciences of Ukraine, Naukova St. 3-b, 79060 Lviv, Ukraine
    \and
    INAF - Osservatorio Astrofisico di Arcetri, Largo E. Fermi 5, I-50125 Firenze, Italy
    \and
    Institute of Physics and Astronomy, University of Potsdam, 14476 Potsdam-Golm, Germany
    \and
    INAF - Osservatorio Astrofisico di Catania, Via Santa Sofia 78, 95123 Catania, Italy 
}

\date{Received ...; accepted ...}

\abstract{When a supernova remnant (SNR) interacts with the dense material of an interstellar cloud, its shock wave decelerates rapidly, and the post-shock temperature drops to levels that permit efficient cooling of the shocked plasma. At this stage, the shock enters the post-adiabatic phase of its evolution. During this phase, the internal structure of the SNR undergoes significant changes, particularly in the immediate post-shock region, at spatial scales relevant to cosmic ray acceleration. Once the shock enters the post-adiabatic regime, the efficiency of diffusive shock acceleration increases due to a higher plasma compression, to a change in the direction of the advection velocity, and to an increased rate of momentum gain. As a result, the momentum spectrum of relativistic particles hardens, deviating from a pure power law at high energies. Particles could reach higher maximum values compared to classical predictions. We highlight the dynamics of post-adiabatic flows in SNRs, study their impact on particle acceleration, and present supporting observational evidence in the radio band.}


\maketitle


\section{Introduction}

\citet{1972ARA&A..10..129W} originally subdivided the evolution of supernova remnants (SNRs) into three phases, followed by a merging with the environment.
Phase I lasts until the swept-up mass becomes comparable to the ejecta mass, then phase II keeps going up to an age comparable to a time scale for the radiative losses close to the shock front, and finally phase III proceeds with the dynamics dominated by the radiative losses. It then appeared that this picture should be supplemented by two intermediate phases (between I-II and between II-III), which have their own distinctive physical features. 

In an updated scheme, but still idealized (e.g.\ valid in the case of a homogeneous ambient medium), the \textit{adiabatic shock} in SNR evolves through three stages: (1) ejecta-dominated or free-expansion \citep{1982ApJ...258..790C,1982ApJ...259..302C}, (2) energy-conversion, from mostly kinetic towards a fixed balance with the thermal one  \citep{2021MNRAS.505..755P, 2017MNRAS.465.3793T}, (3) 'adiabatic' or Sedov-Taylor, with a constant ratio between kinetic and thermal energy \citep{1946JApMM..10..241S,1950RSPSA.201..159T}. 
After the radiative losses of plasma become important, a \textit{radiative shock} develops further through \citep{1988ApJ...334..252C,1992ApJ...392..131S, 1998ApJ...500..342B, 2016MNRAS.456.2343P,2018MNRAS.479.4253P} (4) a partially-radiative or 'post-adiabatic' stage and then (5) a fully-radiative one which is concluded by (6) a dissipation phase. The structure of the flow undergoes essential transformations through the fourth stage, while it remains as simple as a 'snow-plow' over the fifth one. 

We often use the words 'stage' and 'phase' as synonyms. The term 'era', however, implies a broader segment of evolution. For instance, the \textit{adiabatic era} encompasses the entire evolution of the SNR from the explosion up to the point where radiative losses begin to affect the dynamics (that is, stages 1-3). From that time onward, the \textit{radiative era} extends until the complete dissipation of the SNR into the surrounding environment (stages 4-6).

In the present paper, we focus on the radiative era and, more precisely, on its initial, the partially-radiative, stage. The transition of an SNR into it has been studied in a number of papers \citep[see][for a review]{2005JPhSt...9..364P}. 
\citet{1988ApJ...334..252C, 1998ApJ...500..342B} focused on the hydrodynamics (HD) of the SNRs, while \citet{1992ApJ...392..131S,2016MNRAS.456.2343P,2018MNRAS.479.4253P} studied their magneto-hydrodynamical (MHD) properties. The beginning and end of the post-adiabatic phase (stage 4 in the sequence listed above) are marked approximately by the `transition time' $t\rs{tr}$ and the `shell-formation time' $t\rs{sf}$. 
The first one, $t\rs{tr}$, points to the end of the adiabatic regime, namely when the radiative losses become prominent, and the expansion parameter $m$ ($R\propto t^{m}$) deviates from the value $2/5$ that remains constant during the Sedov stage. The second time, $t\rs{sf}$, corresponds to the beginning of the formation of a thin, fully radiative shell outlining the SNR. 
The adiabatic regime of expansion is valid up to $t\rs{tr}$, while the fully radiative one is valid after $t\rs{sf}$. An analytical solution for the radiative shock \citep{1986KFNT....2...15P,2004A&A...419..419B} can be used after $t\rs{sf}$.

One of the important conclusions from the HD studies is that the shock undergoing radiative losses is more compressible than the adiabatic one. In the purely hydrodynamic simulations, the post-shock density can rise to a hundred times the pre-shock value. 
Such a large increase in density can lead to prominent non-thermal emission, due to rise of injected CRs density, even if the efficiency of cosmic ray (CR) acceleration decreases \citep{2015ApJ...806...71L,2024ApJ...974..201D,2025ApJ...980..167D}. 
However, the tangential component of the magnetic field (MF) can significantly alter the spatial structure of the flow downstream and radically limit the density increase downstream \citep{2016MNRAS.456.2343P,2018MNRAS.479.4253P}. This happens because the tangential MF is compressed around the partially radiative shock, together with the matter, which causes the magnetic pressure to grow to a dynamically important level. 
As a consequence, the magnetic pressure 
essentially limit the compression of the radiative plasma compared to the case when MF is negligible. 
In this way, the non-thermal evidence of a radiative shell, as predicted by purely hydrodynamic considerations, will vanish \citep{2025ApJ...980..167D}. 
In contrast, the plasma compression is not influenced by the radial MF component. 

Besides the larger compression, the post-adiabatic flow features another important property. There are rapid and substantial changes in the spatial structure of the plasma velocity $u$, which happen very close to the shock, where the diffusive shock acceleration of particles is provided by the gradient of this velocity, $du/dx$. 

Therefore, in the present paper, we address the CR acceleration at the post-adiabatic shocks by expanding the approach developed by \citet{2024A&A...688A.108P}. The main idea in that study was that the non-uniformity of the downstream flow on the length scales involved in the particle acceleration could affect the CR spectrum. This is a sort of analogy to the non-linear acceleration, where the flow velocity is not spatially uniform upstream due to the back reaction of CRs. In our case, the difference is in the physical reason for this flow non-uniformity: it is a purely hydrodynamical effect taking place downstream of the shock. 

The radiative losses of the plasma are proportional to $n^2\Lambda(T,\tau)$ where $n$ is the density and $\Lambda(T,\tau)$ the cooling coefficient which depends on the temperature $T$ and the ionization time-scale $\tau$. 
Generally, radiative losses become prominent when the shock speed $V$ slows down to values that provide a temperature decrease $T\propto V^2$, which pushes $\Lambda(T)$ to its maximum. However, the post-adiabatic evolution is not a feature of old SNRs only. It is also natural for the SNRs interacting with dense ambient material, like molecular or atomic clouds, since radiative losses are proportional to $n^2$. 
Interacting SNRs are expected to be sources of hadronic gamma-rays. It is worth studying, therefore, what kind of changes in the spectrum of particles accelerated at the post-adiabatic shock one could expect. 

In the present paper, we adopt 1D MHD numerical simulations. As will become clear at the end of the study, CR acceleration at post-adiabatic shocks merits further investigation in higher dimensions. Therefore, the present paper should be considered a first-cut assessment of the problem.

\begin{figure}
  \centering 
  \includegraphics[trim=6 80 10 6,clip,width=\columnwidth]{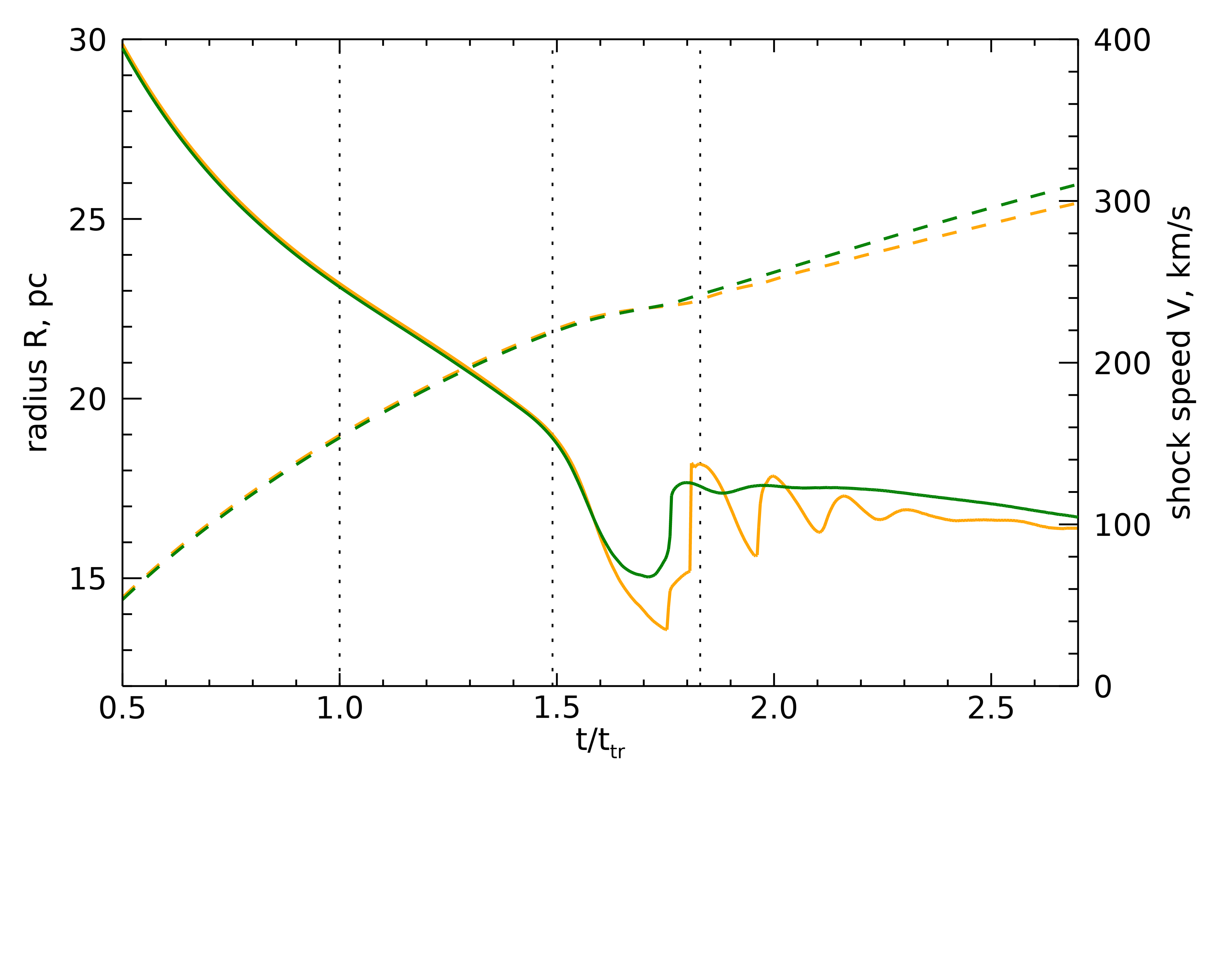}  \includegraphics[trim=6 80 10 0,clip,width=\columnwidth]{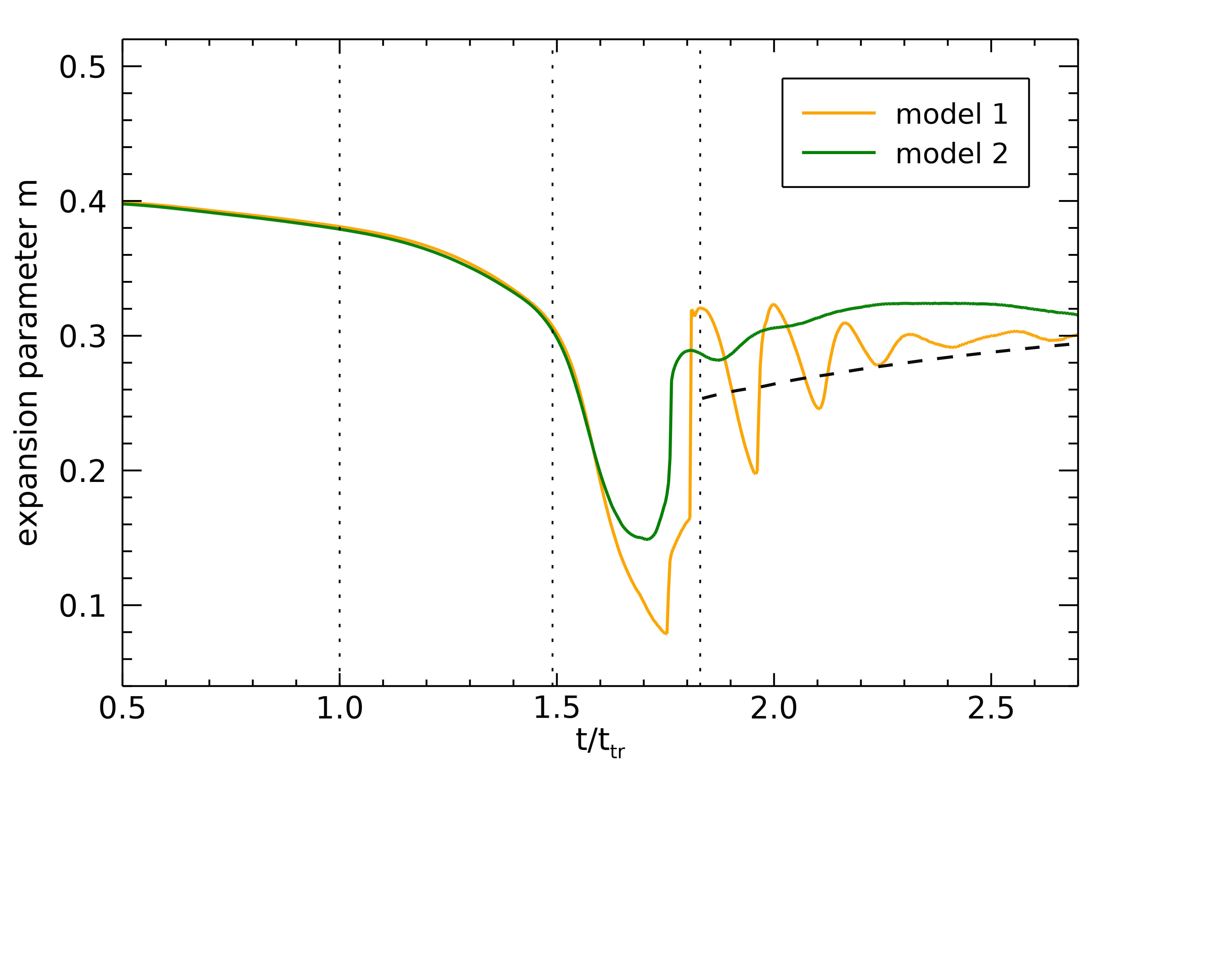}
  \caption{\textit{Top.} The evolution of the radius $R$ (dashed lines) and speed $V$ (solid lines) of the forward shock. \textit{Bottom.} The time dependence of the expansion parameter $m$. Three vertical lines mark the times $t\rs{tr}$ (left), $1.49t\rs{tr}$ (middle) and $t\rs{sf}$ (right).  
  }
  \label{postad:fig-m}
\end{figure}
\begin{figure*}
  \centering 
  \includegraphics[trim=18 100 15 24,clip,width=0.93\columnwidth]{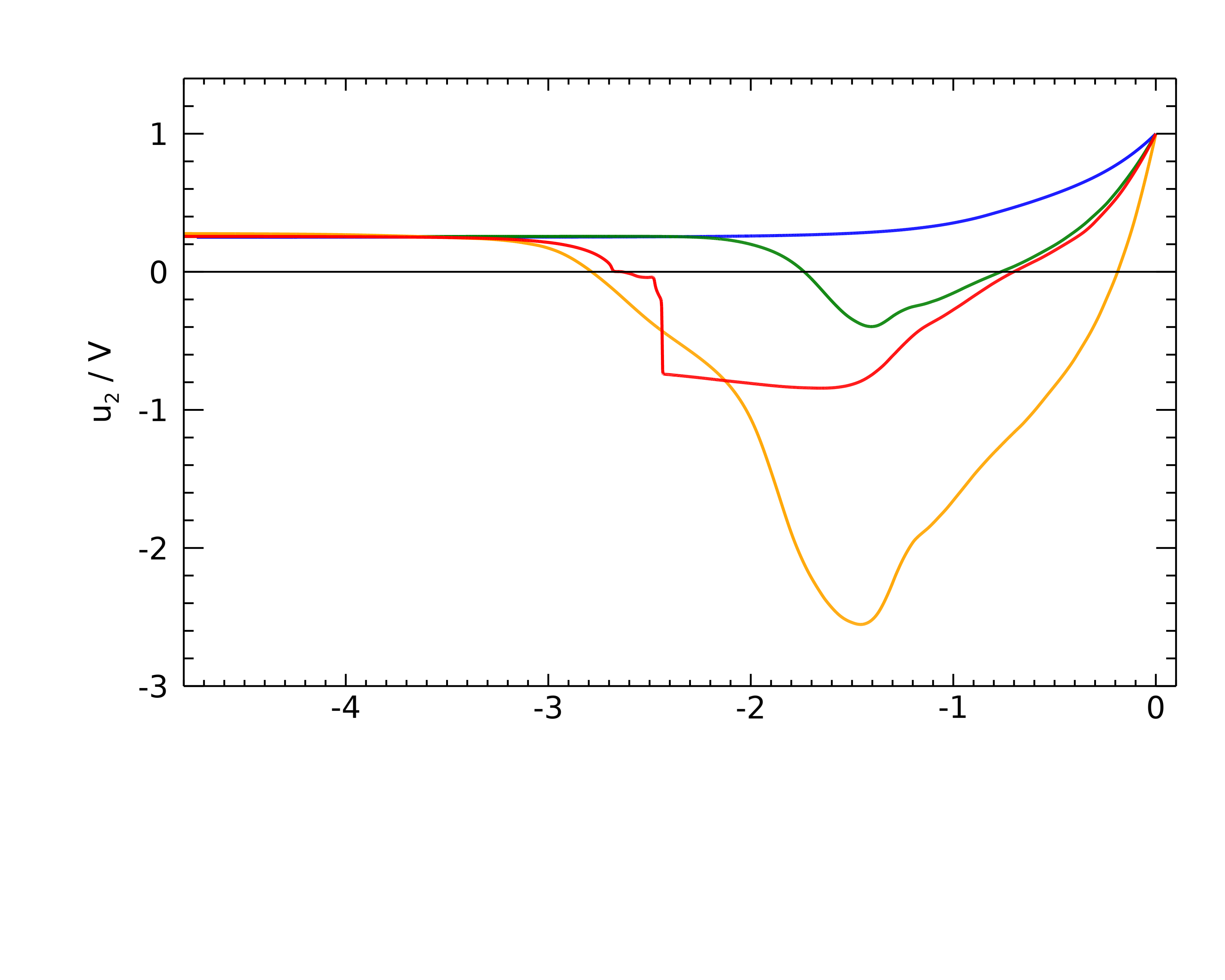}\ 
  \includegraphics[trim=18 100 15 24,clip,width=0.93\columnwidth]{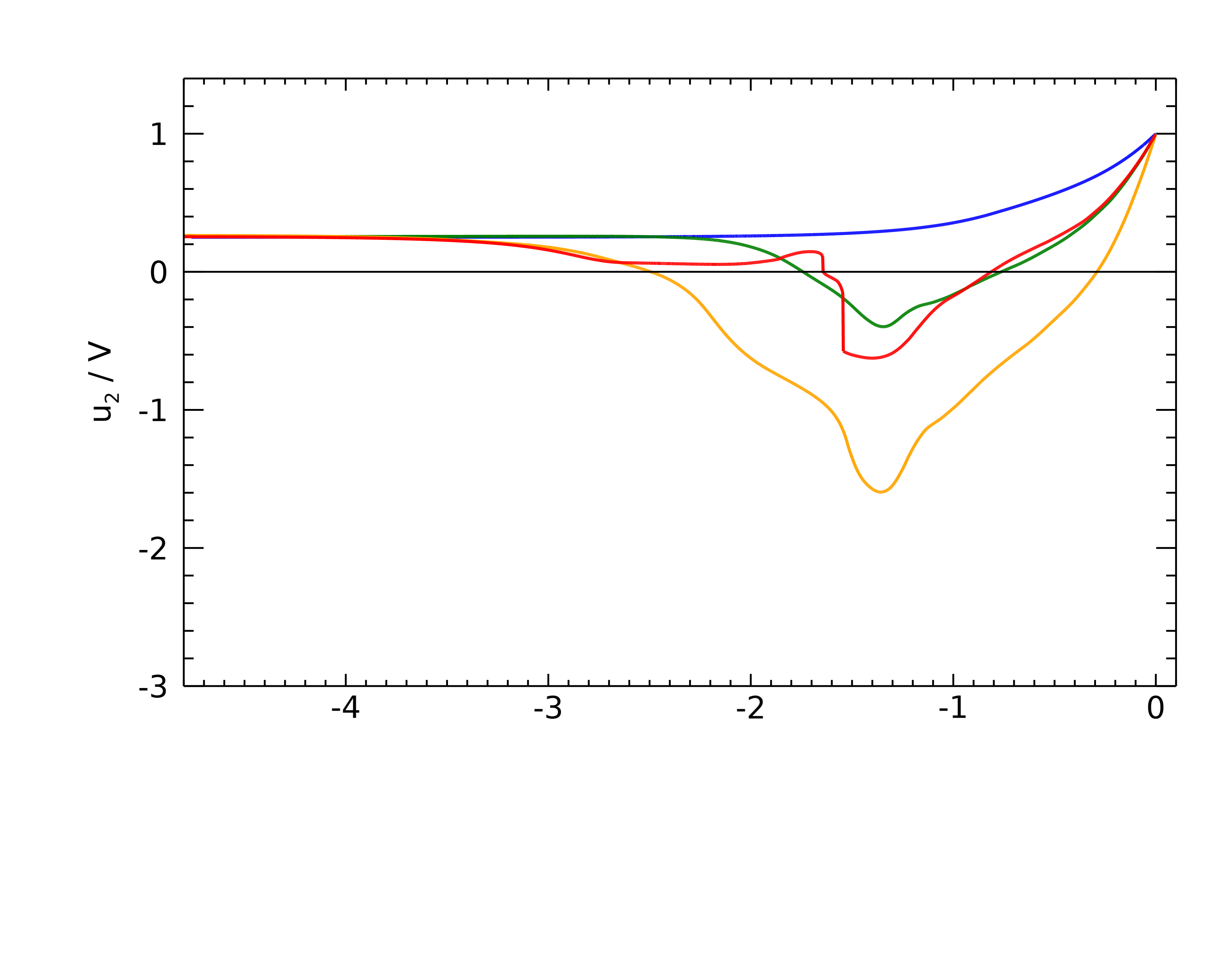}\\
  \includegraphics[trim=18 100 15 24,clip,width=0.93\columnwidth]{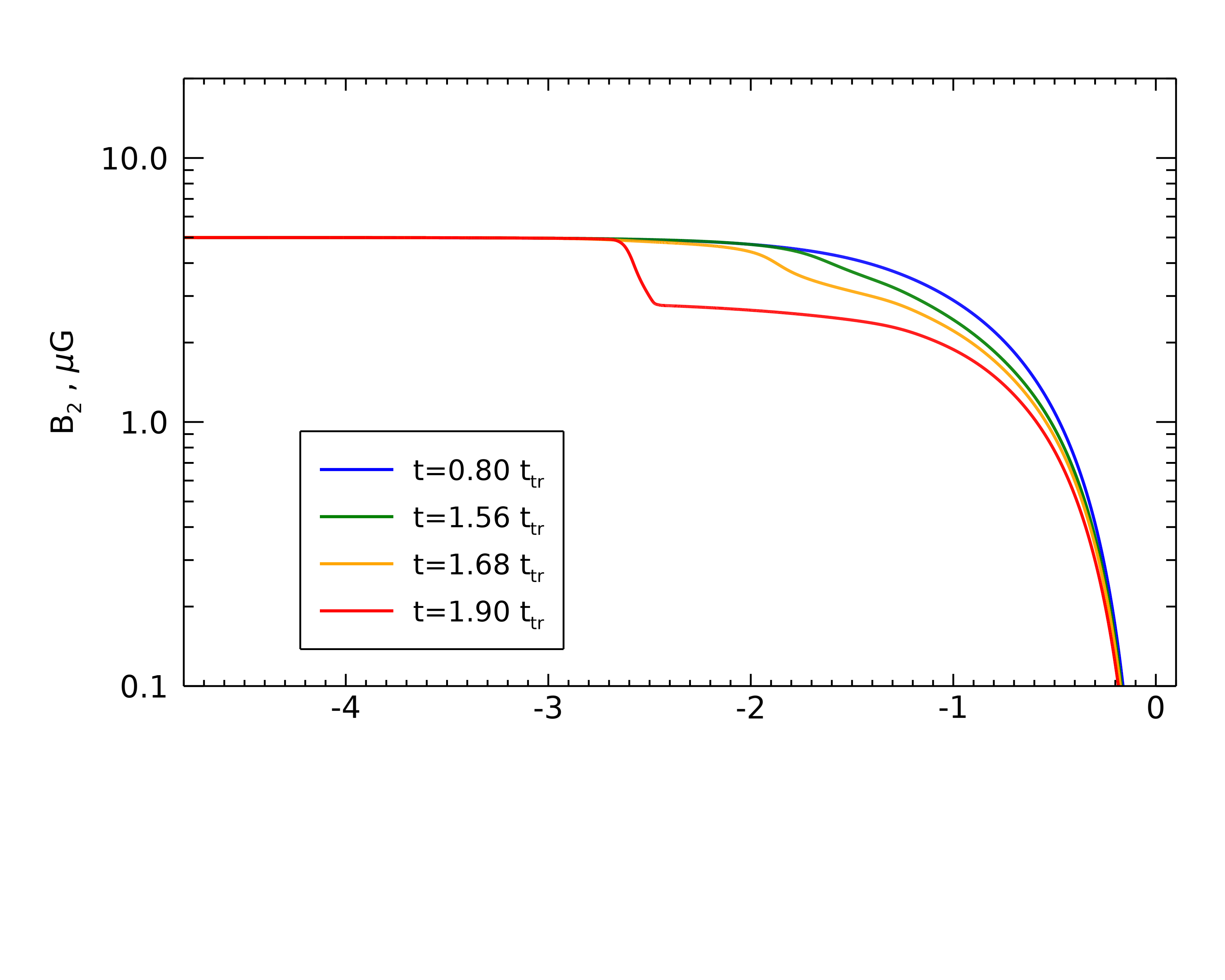}\ 
  \includegraphics[trim=18 100 15 24,clip,width=0.93\columnwidth]{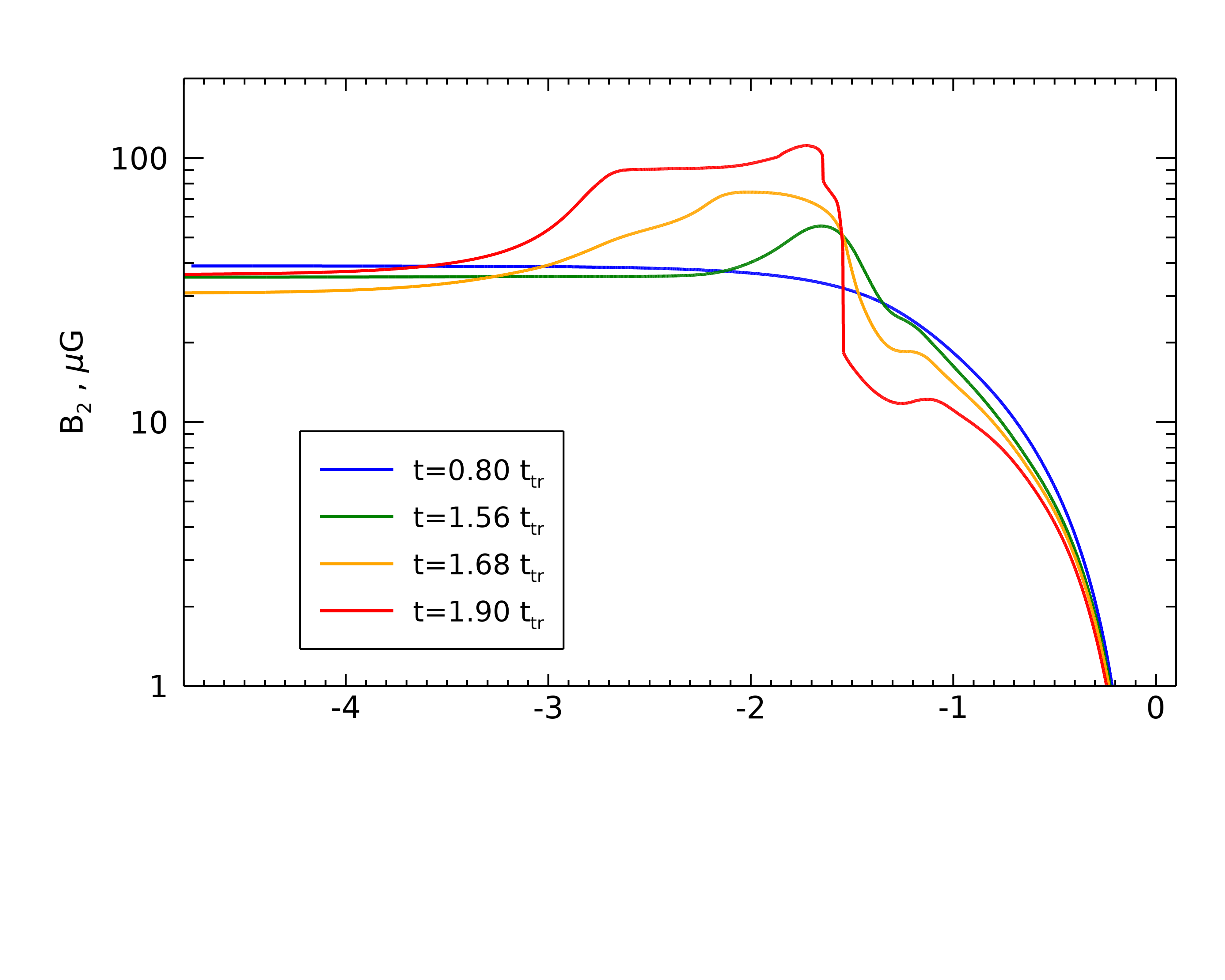}\\
  \includegraphics[trim=18 80 15 24,clip,width=0.93\columnwidth]{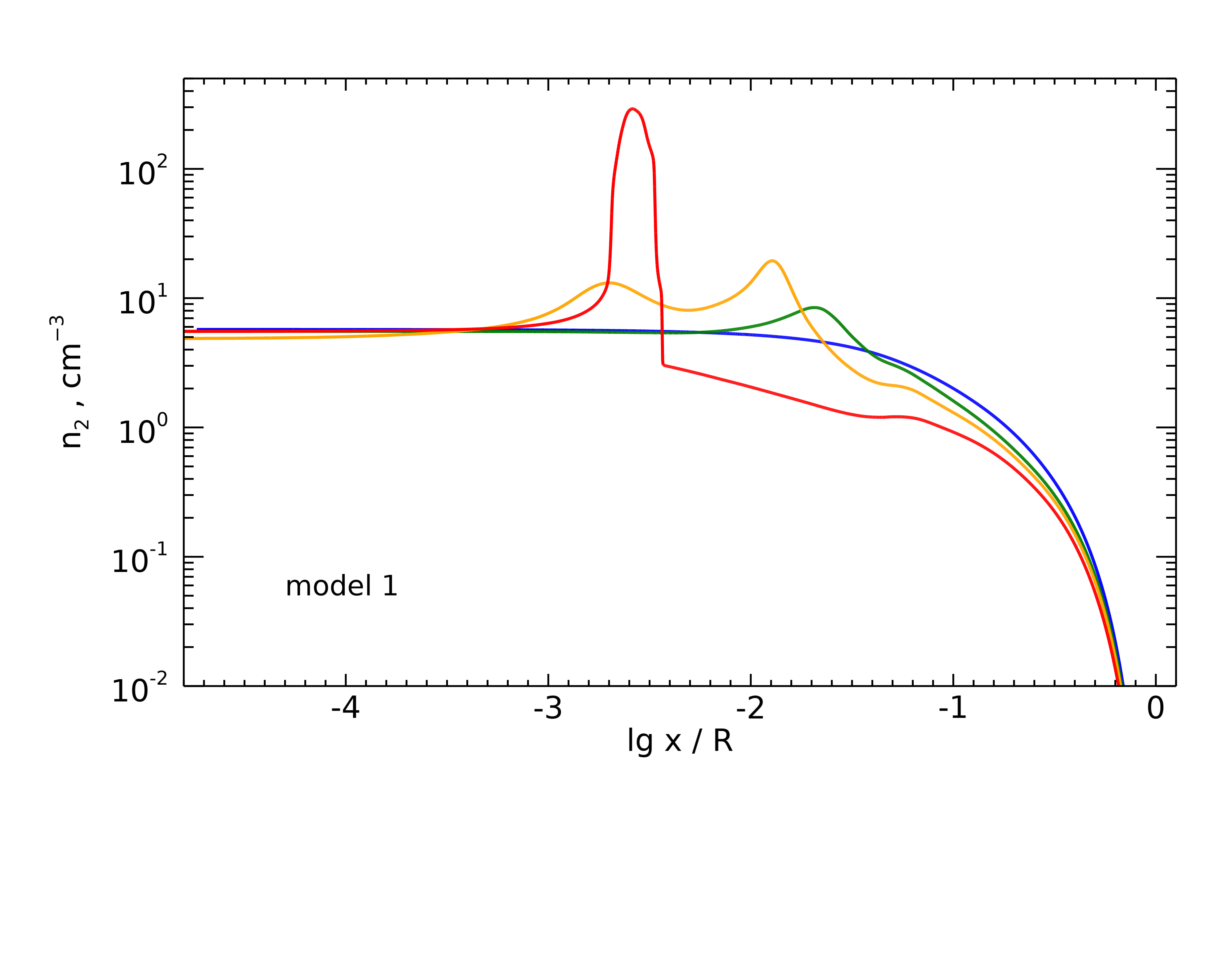}\ 
  \includegraphics[trim=18 80 15 24,clip,width=0.93\columnwidth]{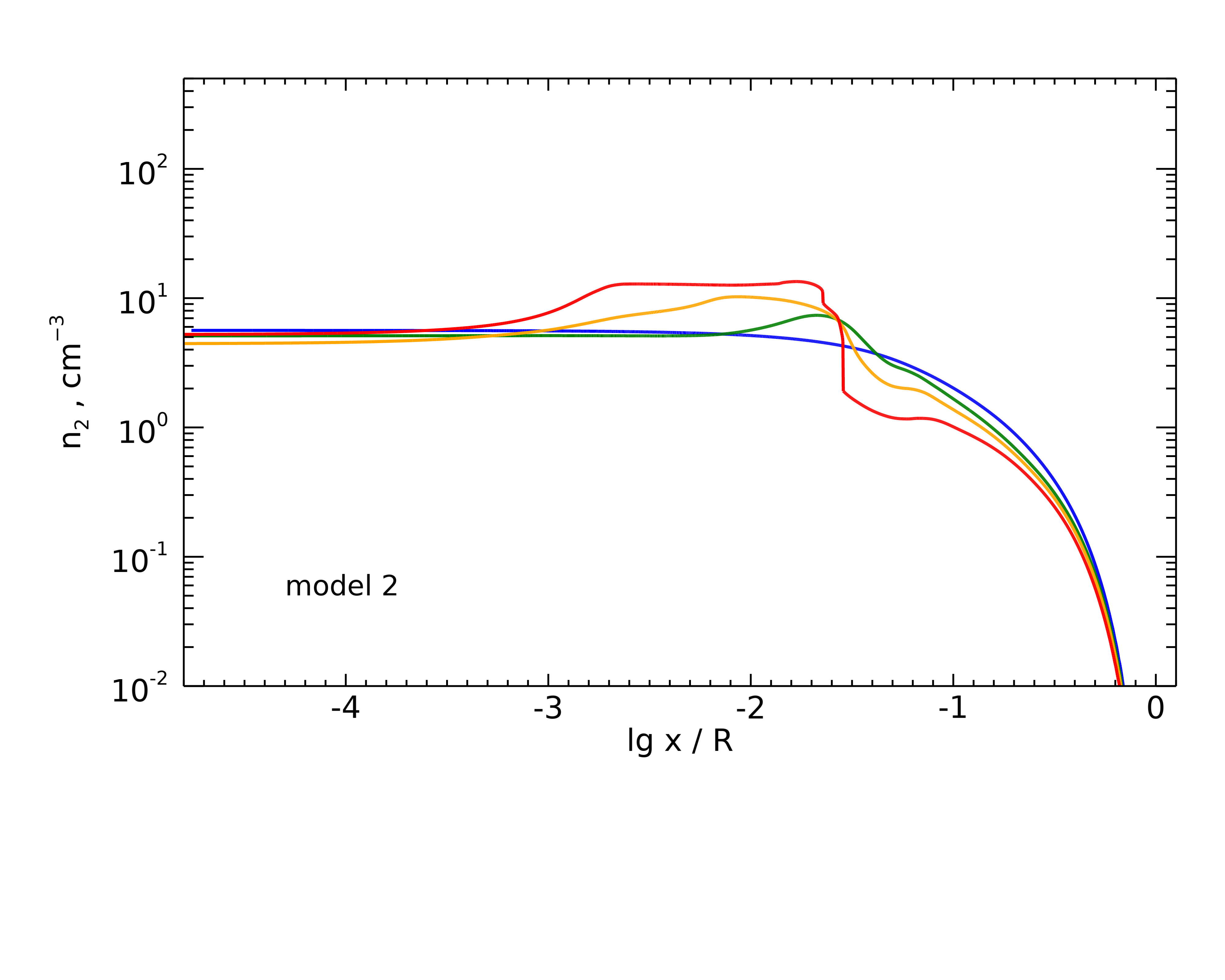}\\
  \caption{Spatial distributions of the flow speed $u_2$, magnetic field strength $B_2$, and number density $n_2$ downstream of the shock at several moments of time. The left column is for the model 1 and the right column for the model 2. The coordinate $x$ marks the distance from the shock, which is used for considerations of the CR acceleration. Note that the horizontal axes are in the log scale. 
  This is necessary to describe equally well variations on both small and large scales relative to the shock position, which are probed by particles with momenta spanning several decades.
  The shock is located on the left of each plot, at $x/R=0$. The SNR center is on the right, at the point with $\lg(x/R)=0$.
  }
  \label{postad:fig-MHDprofiles}
\end{figure*}

\section{Motivations}

\subsection{MHD features of the post-adiabatic SNRs}
\label{postad:MHD-models}

In our previous papers \citep{2016MNRAS.456.2343P,2018MNRAS.479.4253P}, we studied the behavior of the magnetic field in SNRs during the transition phase from adiabatic to fully radiative evolution as well as its role in the modification of the dynamics of post-adiabatic flows. 
In particular, we explored the evolution of radial profiles of density, pressure, and temperature, as well as of the radial and tangential components of MF. In the present paper, we focus instead on the flow velocity profiles. 

We have performed numerical simulations of 1D MHD models in spherical coordinates 
which reflect the evolution of a sector in SNR, with a small solid angle. The initial conditions correspond to the Sedov-Taylor stage. They are provided by setting up a strong point explosion with energy $E\rs{0}=10^{51}\un{erg}$, released isotropically, in the uniform medium with hydrogen number density $n\rs{1}=1\un{cm^{-3}}$ and magnetic field with strength $B\rs{1}$.
We consider two models: \textit{model 1} corresponds to a parallel shock (i.e. ambient MF $\mathbf{B}\rs{1}$ parallel to the shock velocity $\mathbf{V}$) with $B\rs{1}=5\un{\mu G}$, while \textit{model 2} assumes a perpendicular shock with $B\rs{1}=10\un{\mu G}$.
The shock modification due to the back-reaction of CRs was not considered.

The simulations were performed using the PLUTO MHD code \citep{2007ApJS..170..228M} on a static uniform grid divided into 100\,000 computational cells. The grid size was set to $32\un{pc}$. In the numerical setup, we used the Characteristic Tracing time integrator and the HLL Riemann solver. 
The radiative losses of plasma were treated under the assumption of ionization equilibrium.
The setup is similar to that adopted by \citet{2016MNRAS.456.2343P,2018MNRAS.479.4253P}. We refer the reader to these papers for a general overview of the post-adiabatic shock properties in the presence of an ambient magnetic field with different orientations to the shock velocity. 

Fig.~\ref{postad:fig-m} shows, for reference, the time dependence of the shock radius $R$, the shock speed $V$, and the expansion parameter $m$. This parameter is useful in determining the evolutionary stage of a supernova remnant. 
Numerically, $m=0.4$ during the Sedov stage, while in the course of the fully radiative phase, its time evolution (black dashed line in Fig.~\ref{postad:fig-m}) is given by an analytical solution \citep[and references therein]{2004A&A...419..419B}. 

In this figure, the two reference times mark approximate limits of the 'post-adiabatic' stage \citep[for a review of different reference times related to cooling of SNRs, see ][]{2005JPhSt...9..364P}. 
The first one is the 'transition' time, when the flow starts to deviate from the \cite{1959sdmm.book.....S} solution due to the radiative losses of plasma, which become prominent around this time. An analytical estimate for it is given by the expression \citep{1972ApJ...178..159C,1998ApJ...500..342B} 
$t\rs{tr}=2.83\E{4}E\rs{51}^{4/17}n\rs{1}^{-9/17}\un{yrs}$ where $E\rs{51}$ is $E_0$ in $10^{51}\un{erg}$. The thin cold dense shell is formed around the 'shell formation' time \citep{1982ApJ...253..268C,1988ApJ...334..252C}, which is $t\rs{sf}=1.83\,t\rs{tr}$ for our models. From this time on, the shock is fully radiative, i.e., all thermal energy of the newly shocked ISM material is quickly radiated away. Between these two milestones, the shock is partially radiative, with radiative losses becoming increasingly more important in the SNR's dynamics. Radiative losses considerably affect the structure of the flow, particularly the flow speed.  

Animations demonstrating changes in the flow structure in this stage are presented in the Appendix~\ref{postad:app-movie-MHD}. The downstream profiles of the flow speed in the shock reference frame $u_2$, magnetic field strength $B_2$, and plasma density $n_2$ in the two models for a few moments of time are shown in Fig.~\ref{postad:fig-MHDprofiles}. The spatial distributions in this figure are shown versus $\lg(x)$ to demonstrate that changes during the post-adiabatic era also affect very small distances from the forward shock. Our simulations resolve the structures down to $10^{-5}$ of the shock radius for $t\geq t\rs{tr}$. 
The flow speed in the reference frame of the shock is $u_2(x)=V-v_2(R-x)$ where $v_2(r)$ is the flow speed in the laboratory frame, $r$ and $x=R-r$ are the distances from the center and from the shock, respectively.

We see, in agreement with the results presented by \citet{2016MNRAS.456.2343P,2018MNRAS.479.4253P}, that the MF rapidly decreases downstream of the parallel shock, while it increases behind the perpendicular shock (middle panels in Fig.~\ref{postad:fig-MHDprofiles}). Actually, the spatial structure of plasma is not affected by the radial MF because plasma flows along the MF lines without any resistance; this model behaves like a non-magnetic one. Instead, the tangential MF lines limit the plasma compressibility, which increases because of radiative losses. As a result, the plasma density in model 2 is lower than in model 1 (lower panels in Fig.~\ref{postad:fig-MHDprofiles}). 

The most important HD parameter for the present paper is the flow velocity. It varies considerably in space and time when radiative losses come into play (the upper panels on Fig.~\ref{postad:fig-MHDprofiles} and animations in the Appendix~\ref{postad:app-movie-MHD}). Contrary to the previous evolutionary stages, changes in the flow velocity happen on the length scales less than a few per cent of the shock radius, i.e., on the scales involved in the cosmic ray acceleration.

\begin{figure}
  \centering 
  \includegraphics[trim=20 80 10 20,clip,width=\columnwidth]{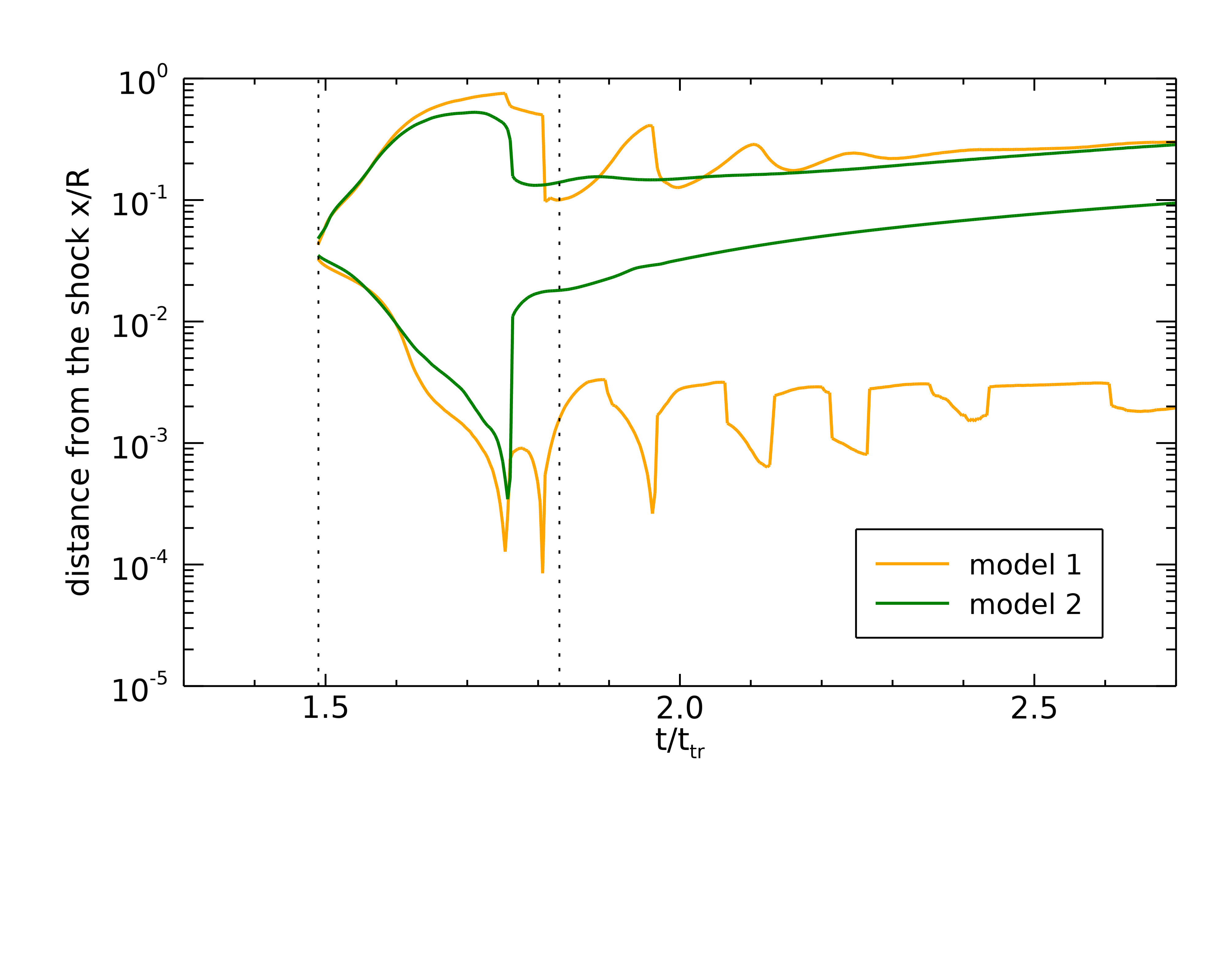}
  \caption{Distances $x\rs{A}$ and $x\rs{B}$ from the shock for the two models. 
  The flow speed $u_2<0$ between the two solid lines. The vertical dotted lines mark the times $1.49t\rs{tr}$ and $t\rs{sf}$. The flow speed $u_2>0$ in the whole domain before $1.49t\rs{tr}$. 
  }
  \label{postad:fig-xab}
\end{figure}

The post-adiabatic flow has a downstream region, between the coordinates $x\rs{A}$ and $x\rs{B}$, where the flow velocity is inverted, $u_2<0$. This happens because the pressure drops right after the shock due to radiative losses of plasma there, while the deeper interior remains hot. In the presence of the tangential component of MF, the magnetic pressure increases in this region due to plasma compression, and the magnetic pressure diminishes the overall effect, that is, the flow speed (Fig.~\ref{postad:fig-MHDprofiles}) and the size of the region with negative $u_2$ (Fig.~\ref{postad:fig-xab}). 

The region with the inverted flow velocity appears first around time $1.49 t\rs{tr}$ and extends almost to the end of the post-adiabatic phase, up to $1.75t\rs{tr}=0.97t\rs{sf}$. Thus, the highest boost in the CR acceleration efficiency should happen around this period.

In summary, there are several prominent features in the shock and flow velocity that can affect the particle acceleration at the post-adiabatic shocks.
\begin{itemize}
\item The shock slows down ($u_1$ diminishes) faster than during the previous adiabatic phase. 
\item The ratio $u_1/u_2(x)$ increases in a certain range of $x$ as radiative losses start to influence the shock. 
\item This ratio may become quite large as $u_2$ approaches zero.
\item During some time interval, there is a region where $u_2$ becomes negative, and particles are not advected away from the forward shock, but instead are pushed back towards it.
\item The spatial extent of this region could reach about $10\%$ of the SNR's radius. 
\item Changes in the spatial structure of $u_2$ occur even at distances $\sim 10^{-3}R$, that is, so close to the shock that they even influence the acceleration of the radio-emitting electrons (as we will see later). 
\item The MF component, orthogonal to the flow velocity, can affect the velocity profile.
\end{itemize}
Only the first property leads to a reduction in the acceleration efficiency; all others increase it. The shape of the CR spectrum and the maximum energy they can attain are determined by a combination of these trends.

\begin{figure}
  \centering 
  \includegraphics[width=\columnwidth]{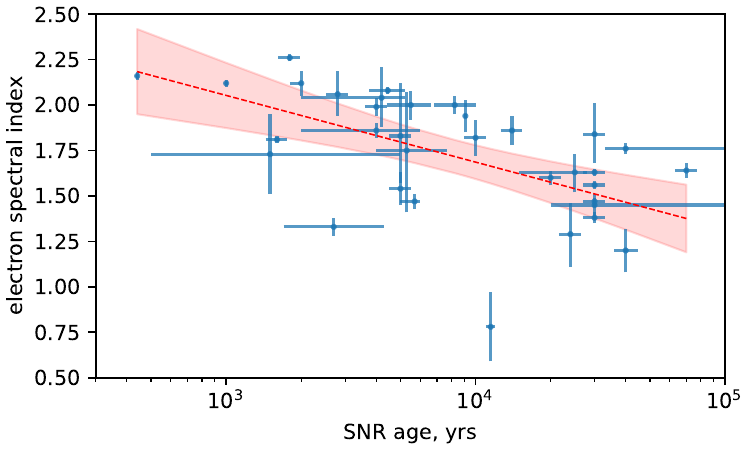}
  \caption{The electron spectral index versus the age for a sample of SNRs with the radio and GeV $\gamma$-ray emission detected, from the data given by \citet{2019ApJ...874...50Z}. 
  The red dashed line represents a linear fit between the index and logarithm of age, with the slope 
  $k = -0.37 \pm 0.05$. The red area shows the 1-$\sigma$ uncertainty. 
  Note about treatment of the data. We excluded MSH~15-56 from the sample because its radio emission is likely dominated by a pulsar. 
  There are few SNRs with two values of $\alpha$ estimated in this reference.  
  We calculated the average $\alpha$ from the two values if the uncertainty is due to various estimates of the ambient density (Tycho, G166.0+4.3, S147, CTB~73B). 
  We take only one value for W51C, RX~J0852-4622, namely that which corresponds to the same spectral model as used for other SNRs.
  }
  \label{postad:fig-zang}
\end{figure}
\begin{figure*}
  \centering 
  \includegraphics[trim=0 1 0 0,clip,width=\textwidth]{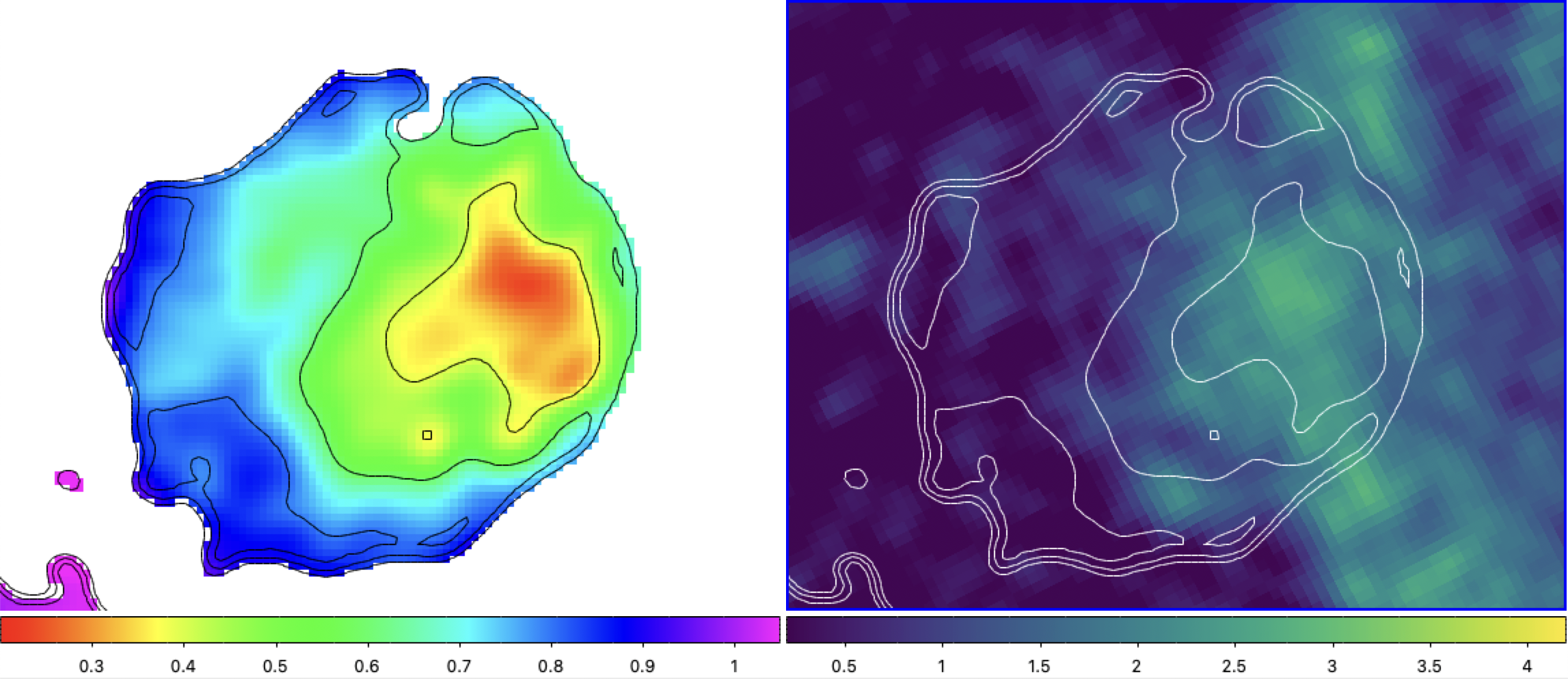}
  \caption{The distribution of the radio spectral index over Kes~73 measured between 1.4 and 5.0 GHz (left) and the 
   ${}^{12}$CO ($J=3\rightarrow2$) intensity map \citep{2013ApJS..209....8D} integrated over the velocity range between 94 and 95 $\un{km/s}$, in units $\un{K\ km/s}$ (right). Contours correspond to the values of the radio index $0.8$, $0.6$, and $0.4$. 
   Also \citet{2017ApJ...851...37L} present ${}^{12}$CO ($J=1\rightarrow 0$) intensity maps around Kes~73 at various velocities. Those in the range 86-97 km/s correlate with the radio index map as well. 
  }
  \label{postad:fig-kes73}
\end{figure*}

\subsection{Observational evidence}
\label{postad:obs-evid}

Simulations show that the flow velocity profile changes continuously after the end of the adiabatic era, as well as the conditions for particle acceleration around the forward shock, which is progressively affected by the increasing radiative losses. Are there observational evidences supporting this? 

What kind of dependence should we look for? The ratio of the pre- to the post-shock velocities $\sigma=u_1/u_2$ increases during some time after $t\rs{tr}$. This occurs in regions very close to the shock and could affect the acceleration of the radio-emitting electrons. If so, then the radio index $\alpha=(s-1)/2$ should reflect changes in the electron spectral index $s=(\sigma+2)/(\sigma-1)$. That is, the absolute value of $\alpha$ should decrease as the SNR progresses into the post-adiabatic phase.
With this trend in mind, we may consider the following two observational hints. 

First, let us consider a sample of SNRs that have radio and GeV $\gamma$-ray emission detected, as well as estimations for their age. \citet{2019ApJ...874...50Z} have analyzed the multi-wavelength spectra of 35 such SNRs under a uniform approach. 
The authors provide estimates for a spectral index and SNR age.
These are SNRs likely interacting with dense clouds, so they are rather in the post-adiabatic stage, with the age $\gtrsim t\rs{tr}$.
A broadband (from radio through X-rays to $\gamma$-rays) spectral fit provides the value of the particle spectral index. The electron spectral indices of these SNRs estimated in this reference are shown in Fig.~\ref{postad:fig-zang} as a function of the SNR age. 
A systematic decline in the index with age is apparent.
It means that, independently of local conditions, the ratio $u_1/u_2$ probed by the radio-emitting electrons increases over time for older SNRs, which could suffer (to a different extent) from radiative losses. Thus, this trend agrees with our hypothesis.

Another possibility is to look for a single SNR where a fraction of a shock interacts with a dense environment. If different parts of the shock enter regions with different densities, then we may expect a lower value of $\alpha$ at locations where the density is higher. There, the shock is affected by radiative losses to a larger extent because the losses are proportional to $n^2$. 
Impressively, the correlation between the radio spectral index and the density is apparent in the supernova remnant Kes~73 (G27.4+0.0).
The left plot in Fig.~\ref{postad:fig-kes73} shows maps of the radio spectral index of Kes~73 obtained from the radio observations on the Green Bank Telescope and the Very Large Array \citep{2014MNRAS.445.4507I}. The CO intensity map in the region around Kes~73 is shown in the right plot. We used CO High-Resolution Survey data \citep{2013ApJS..209....8D}\footnote{Available at \url{http://dx.doi.org/10.11570/13.0002}.} to produce this map. 
It is important to note that this SNR is young and, as a whole, is not in the partially radiative phase, let alone in the fully radiative phase  \citep{2017ApJ...846...13B}. We believe that only a small fraction of the forward shock is encountering a molecular cloud, and radiative effects are initiating their development in these isolated regions. 
Some aspects of the hypothesis that the variation of the radio spectral index in Kes~73 could be due to the shock interaction with dense cloud are discussed in Appendix~\ref{postad:appkes73}.

\section{CR acceleration at a post-adiabatic shock}
\label{postad:sect-CRacc}

\subsection{General considerations}

Let us consider two reference frames. The first is the observer’s frame, which is at rest relative to the explosion center. In this frame, the shock velocity $V$ in the 1D sector of the SNR is positive. The second is the shock frame, often used in the description of CR acceleration. In this frame, the shock is at rest, and the positive direction of the coordinate $x$ is defined along the flow velocity of the upstream medium.

What could happen with cosmic ray acceleration at the shock when, for some time, material in some region downstream has, in the frame of the observer, a radial velocity higher than the shock velocity itself? 
Under such conditions, 
this translates into `negative' velocities of this material in the reference frame comoving with the shock.

Such a negative velocity in the shock reference frame means that the plasma moves toward the shock in the downstream region, not out of it (as commonly assumed). Such a situation may naturally occur in SNRs right after the end of the Sedov phase when the radiative losses of plasma in the vicinity of the shock become prominent. 

As a preliminary point, we have to keep in mind that in the ``standard'' model for diffusive particle acceleration, in the shock reference frame, the flow velocity is always taken to be positive, namely the medium moves towards the shock surface in the upstream flow, while it moves away from it downstream. This has a clear consequence in the evolution of particles, as depicted, for instance, in Bell's model for particle acceleration in shocks. Particles can be accelerated as long as they are hopping back and forth across the shock. The ingredients of the model are the following \citep{1991SSRv...58..259J}:\\
1. the average increase in momentum experienced by a particle over a single cycle; it is $\Delta p/p \simeq 4/3\,(u_1-u_2)/c$, where $(u_1-u_2)$ is the velocity difference between the upstream (labeled with 1) and the downstream (labeled with 2) flow;\\
2. the probability that a particle in the downstream region can reach the shock for another cycle, traveling against the general downstream flow.

In this model, 
quantities that can change and that depend both on the ambient magnetic field and the particle momentum are: (i) the maximum distance from the shock a particle can reach on the shock upstream $x\rs{p}\sim D/u$, where $D$ is the diffusion coefficient; the same expression is also valid for the maximum distance it can reach downstream, under the condition that it still has some chance to return to the shock again; 
say $D=\eta\,r_{\rm g}c/3$, where $r_{\rm g}=r_L\gamma=p\,c/(e\,B)$ is the radius of gyration ($r_L$ being the Larmor radius) and $\eta\ge1$; (ii) the residence time on each side is nearly the previous quantity divided by $c$, so the time required for the particle to complete one full cycle is $t_{\rm cycle}=(4/c)(D_1/u_1+D_2/u_2)$.

\begin{figure}
  \centering 
  \includegraphics[trim=230 300 210 80,clip,width=\columnwidth]{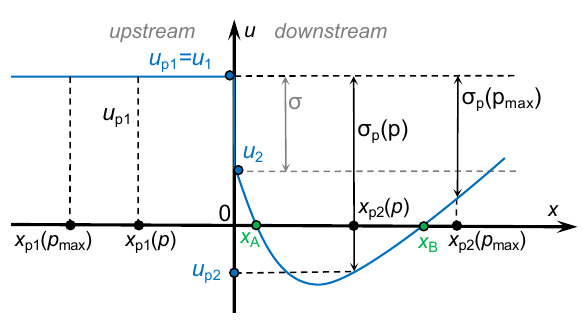}
  \caption{Diagram illustrating the spatial structure of the flow velocity (shown with the blue line), in the reference frame of the shock, which is located at $x=0$. 
  The diffusion length of a particle with momentum $p$ is $x\rs{p1}$ upstream and $x\rs{p2}$ downstream. These are the largest distances the particle can diffuse and return to the shock to be further accelerated. The shock compression factor $\sigma=u_1/u_2$ is the ratio of the immediate pre- and post-shock velocities. Instead, $\sigma\rs{p}(p)=u\rs{p1}(p)/u\rs{p2}(p)$ is an effective compression factor `seen' by a relativistic particle with momentum $p$; it varies with momentum. The shock here is not CR-modified and, therefore, $u\rs{p1}(p)=u_1$ upstream, for any momentum. For the downstream region, one may use an approximation $u\rs{p2}(p)\approx u_2(x\rs{p2})$, see Sect.~\ref{postad:sect-f0}.
  }
  \label{postad:fig-sxema}
\end{figure}

Now, this framework requires that particles may eventually flow away in the downstream flow; otherwise we will have a runaway situation, in which no particle can escape, but they all are accelerated `forever'. Where is the trick? The standard model of diffusive particles acceleration assumes steady-state conditions; while a situation with a negative $u_2$ and a shock strong enough to efficiently accelerate particles cannot continue forever, otherwise we end up with an infinite density in the vicinity of the shock and infinite energy in the accelerated particles.

Let us consider a strong non-relativistic shock discontinuity. In units of the shock speed, the immediate downstream velocity after the shock will be $u_2(\hbox{\rm immediate})=u_1/\sigma$ with the compression factor $\sigma=4$. Let us then assume a downstream velocity profile in which, at a given time, the downstream velocities turn negative from the distance $x\rs{A}$ to the distance $x\rs{B}$ from the shock. We recall that, in the shock frame, the shock is at $x=0$, $x\rs{A}$ and $x\rs{B}$ are positive and $x\rs{A}<x\rs{B}$ (Fig.~\ref{postad:fig-sxema}). The diffusion length $x\rs{p2}$ could be less than $x\rs{A}$, in between these two coordinates, or larger than $x\rs{B}$, depending on momentum.

For the particles, as soon as their diffusion length $x\rs{p}<x\rs{A}$, essentially nothing changes from the common picture of acceleration because during their acceleration, they do not notice a velocity inversion further on. In particular, only a fraction of particles can return to the shock against the flow, specifically those with an $x$-component of their velocity $w$ at a given point $x$ satisfying $w_x < -u_2(x)$. Instead, particles in the region of the negative $u_2$ will be easily directed toward the shock by the flow. Only particles with $w_x > -u_2(x)$ can escape this region toward the far downstream.

The idea is that, when the accelerated particles gain the momentum $p_A$ corresponding to the diffusion length $x\rs{p}=x\rs{A}$, they would remain trapped between the shock and the region with negative $u$, until they reach the value $p_B$ relevant to $x\rs{p}=x\rs{B}$ that allows them to be advected downstream or otherwise, if the hydrodynamic evolution proceeded faster than the acceleration process, until the region of negative $u$ eventually disappears. 
Since generally $x_B$ is a non-negligible fraction of the SNR radius (Fig.~\ref{postad:fig-xab}), this means that particles, if trapped long enough, could be accelerated up to very high energies.

\begin{figure}
  \centering 
  \includegraphics[trim=30 80 15 15,clip,width=0.92\columnwidth]{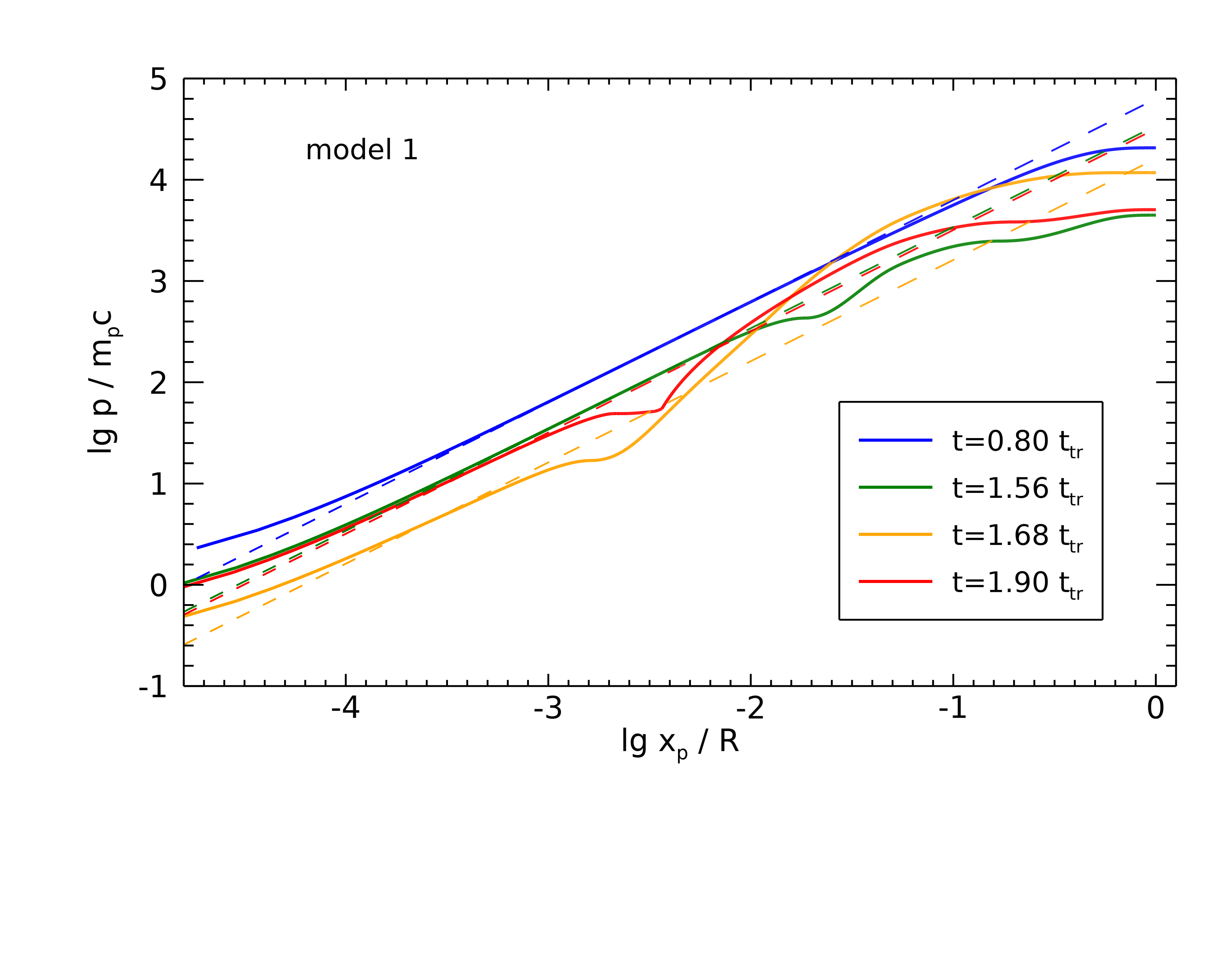}
  \includegraphics[trim=30 80 15 15,clip,width=0.92\columnwidth]{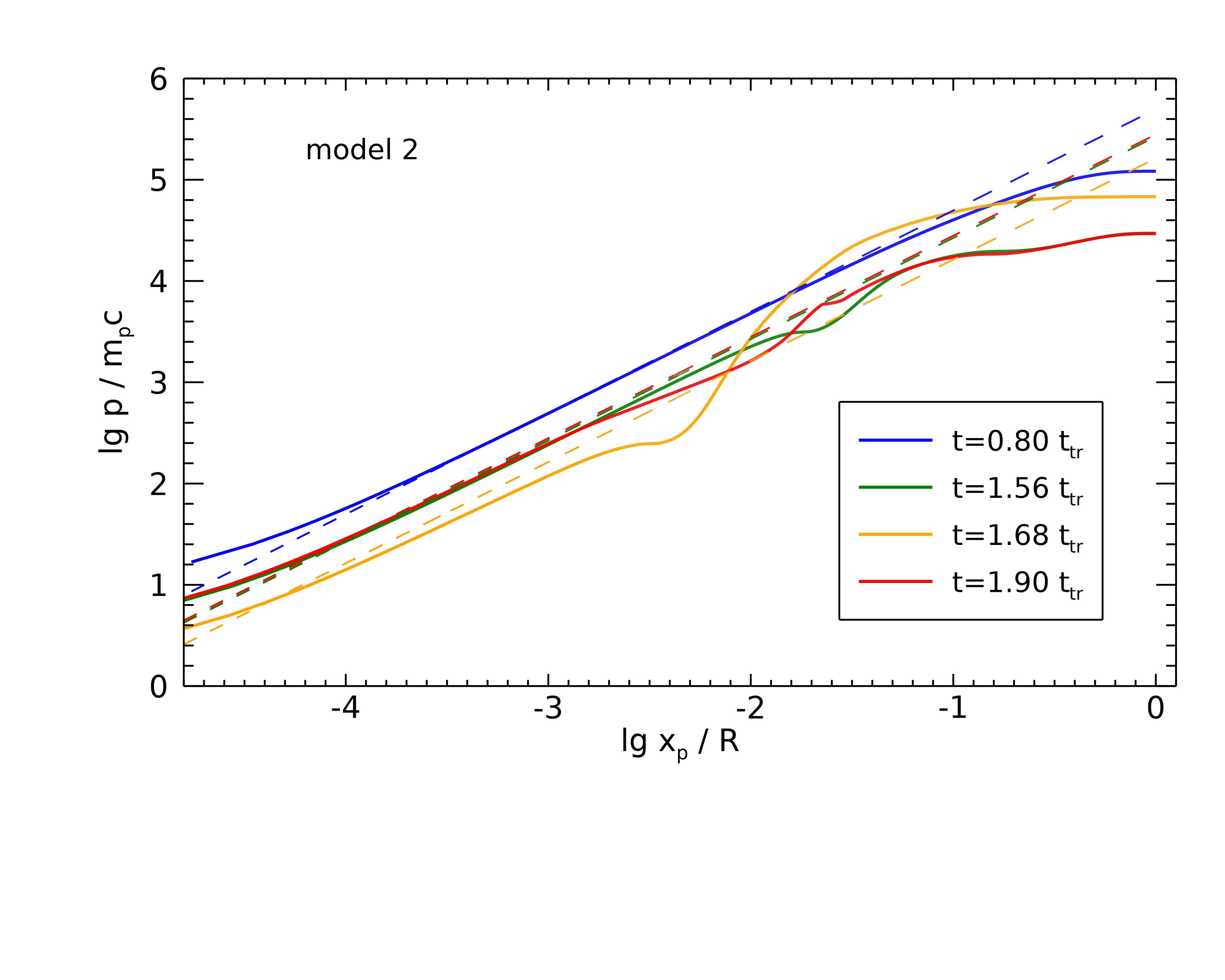}
  \caption{The minimum momentum $p$ of particles which may reach the distance $x\rs{p}$ downstream and return to the shock to continue acceleration. It is calculated from the equation $p=\bar p(x\rs{p})$ for Bohm diffusion. Lines with different colors correspond to a few time moments. Dashed lines show $p$ calculated in a common approach, from $x\rs{p}=D_2(p)/u_2$ with uniform $u_2$ and uniform $B_2=5\un{\mu G}$. \textit{Top} plot for the model 1 and \textit{bottom} plot for the model 2.
  }
  \label{postad:fig-Fxx-appr0}
\end{figure}

\subsection{Diffusion length}

The question is how to compute more precisely $x\rs{p}$, considering the spatial variation of hydrodynamic quantities, like $u$ and $B$. In the case with constant profiles of $D(x)$ and $u(x)>0$, the diffusion length $x\rs{p}$ of particles with a given $p$ is such that $ux\rs{p}/D = 1$. One could impose a generalization to non-constant profiles by using the simplest dimensional criterion as 
\begin{equation}
 \int_0^{x\rs{p}} \frac{|u(x')|}{D(x')}\,dx'=1
 \label{postad:def-xp-appr0}
\end{equation}
where the absolute value of velocity is taken because $x\rs{p}$, $dx'$, and $D(x')$ are always positive while $u(x')$ can change the sign. 

We consider the diffusion coefficient $D=\eta cr_0\,(r\rs{g}/r_0)^{1/\kappa}/3$, where $r_0$ is a reference length, and $\kappa=1,2,3$ for the Bohm, Kraichnan, and Kolmogorov diffusion, respectively. By substituting equation (\ref{postad:def-xp-appr0}) for $x\rs{p}$ with this diffusion coefficient, we arrive at the function $\bar p(x\rs{p})$ which relates the momentum $p$ to a given diffusion length $x\rs{p}$ 
\begin{equation}
 \bar p(x\rs{p})=\left(\frac{3}{\eta c r_0}\right)^\kappa
 \frac{er_0}{c}
 \left[\int_0^{x\rs{p}}|u(x)||B(x)|^{1/\kappa} dx\right]^\kappa.
 \label{postad:eq-pFxx-appr0}
\end{equation} 
It is important that the derivative is always $d\bar p/dx\rs{p}\geq 0$. Therefore, there is no multiplicity in $x\rs{p}$ for any momentum at any time.

Fig.~\ref{postad:fig-Fxx-appr0} shows the momentum $p$ as a function of the downstream diffusion distance $x\rs{p}$ in our MHD models. It is calculated as a solution of the equation $p=\bar p(x\rs{p})$ with the function $\bar p(x)$ given by  Eq.~(\ref{postad:eq-pFxx-appr0}). Particles with $(x,p)$ below the lines cannot return to the shock for acceleration. 

For the sake of comparison, we also plot the dashed lines that correspond to the common approach, with a uniform flow downstream. 
We see from this figure (cf. Fig.~\ref{postad:fig-MHDprofiles}) that the non-uniformity of $u_2(x)$ and $B_2(x)$ may affect particles with quite small diffusion length, down to $\sim 10^{-3}$ of the shock radius, for the Bohm diffusion. These are the particles with momenta as low as $\sim 10\un{GeV}$ in model 1 (yellow line in Fig.~\ref{postad:fig-Fxx-appr0}), i.e., those emitting radio waves. If the Bohm factor $\eta$ is considerably larger than unity or the Kolmogorov diffusion operates \citep[as observed, ][]{2018MNRAS.480.2200S,2025ApJ...994..147P} then the diffusion coefficient and therefore $x\rs{p}$ are larger for the same momenta. Then, the radio-emitting electrons could probe larger distances from the shock, at least some time during the post-adiabatic phase.

The flow velocity profile changes over time with a stronger spatial variation around $1.7t\rs{tr}$ (yellow lines in Fig.~\ref{postad:fig-Fxx-appr0}) when it affects particles with the smallest momenta.
At different times and different distances, particles probe different compression factors $u_1/u_2(x\rs{p})$ and get different $\Delta p$ that should be reflected in their momentum distribution and in the radio spectrum they emit.

\subsection{Increment in momentum}

Another aspect of particle acceleration at the post-adiabatic shocks is related to the momentum gain per cycle. Let us consider the interaction of a particle (mass $m$, speed $w\sim c$) with scattering centers (mass $M\gg m$ with velocities $u_1>0$ upstream and $u_2>0$ downstream, spatially constant in their domains and with absolute values $u_1\ll w$, $u_2\ll w$). 
The initial velocity of a particle is $w\rs{o}=-w<0$ toward a scattering center in the upstream flow. After interaction there, its velocity is $w'=w+\mu u_1$ (to the order $u_1/w$) where $0\le\mu\le1$ is the cosine of the angle between the particle and the flow velocities. Then the particle crosses the shock, interacts with a scattering center in the downstream flow, and its velocity becomes $w''=-(w'-\mu u_2)$. Therefore, the increment in momentum for a given particle in one acceleration cycle is $\Delta p/p=(w''-w\rs{o})/w\rs{o}=\mu (u_1-u_2)/w$. 
After averaging over the particle flux proportional to $w\mu$, we arrive at a known formula $\Delta p/p \simeq 4/3\,(u_1-u_2)/c$. 
Instead, if $u_2<0$, then $w''=-(w'+\mu |u_2|)$ and $\Delta p/p=\mu (u_1+|u_2|)/w$; and the flux average yields $\Delta p/p \simeq 4/3\,(u_1+|u_2|)/c$.

During one cycle, particles with momentum $p$ experience many collisions, but the whole process proceeds in a way that the overall momentum increase during a cycle is determined primarily by the outermost interactions, at the locations $x\rs{p1}$ and $x\rs{p2}$ where the flow velocities are $u\rs{p1}$ and $u\rs{p2}$ (see Fig.~\ref{postad:fig-sxema}).
Indeed, it seems that the velocity gains and losses in intermediate interactions between these two points, on average, nearly cancel themselves out. 
Such a property is demonstrated by a solution for the CRs-modified shock where the spectral index at the momentum $p$ may be estimated from the effective compression $u_1(x\rs{p1})/u_2$. 
Therefore, by introducing $\sigma\rs{p}=u\rs{p1}/|u\rs{p2}|$, the general formula becomes
\begin{equation}
 \frac{\Delta p}{p}=\frac{4}{3} \frac{u\rs{p1}}{v}\frac{\sigma\rs{p}-\mathrm{sgn}(u\rs{p2})}{\sigma\rs{p}}
\end{equation} 
where $\mathrm{sgn}(x)$ is the sign function. 
We can see the two effects from this expression: the first is due to the negative $u_2$ and the second is due to the change in the effective compression $\sigma\rs{p}$, which may be quite high if $u\rs{p2}$ approaches zero. 
The factor $(\sigma\rs{p}-\mathrm{sgn}(u\rs{p2}))/\sigma\rs{p}$ increases linearly with the decrease of the ratio $u\rs{p2}/u\rs{p1}$, e.g., from 0.75 for $u\rs{p2}/u\rs{p1}=0.25$ to 1.25 for $u\rs{p2}/u\rs{p1}=-0.25$. 
This means that if there is a downstream region of the negative flow velocity, the increase in momentum per acceleration cycle could be about two times the classic value (for uniform and positive $u_2(x)$). 

In contrast, the shock decelerates considerably over time during the post-adiabatic phase. Therefore, the temporal variation of $\Delta p/p$ is determined by changes in the HD structure of the flow as well as in the shock speed.

\subsection{Spectrum of accelerated particles}
\label{postad:sect-f0}

What about the particle distribution function? 
The solution of the steady-state kinetic equation for CRs accelerating in the spatially variable flow downstream has been found by \citet{2024A&A...688A.108P}. 
The distribution function is 
\begin{equation}
 f\rs{o}(p)=\frac{\xi n_1}{4\pi p\rs{o}^3}\frac{3u_1}{u\rs{p1}-u\rs{p2}}
 \exp\left[-\int_{p\rs{o}}^{p}\frac{3u\rs{p'1}}{u\rs{p'1}-u\rs{p'2}}\frac{dp'}{p'}\right] 
 \label{postad:f0sol}
\end{equation}
where $\xi$ is the injection efficiency. 
The spatial variation of the flow velocity in the downstream region is accounted for through $u\rs{p2}$ and through $u\rs{p1}$ in upstream locations.
\begin{equation}
 u\rs{p1}(p)=u_1-\int_{-\infty}^{-0}\frac{f(x,p)}{f\rs{o}(p)}\frac{du}{dx}dx,
\end{equation}
\begin{equation}
 u\rs{p2}(p)=u_2+\int_{+0}^{+\infty}\frac{f(x,p)}{f\rs{o}(p)}\frac{du}{dx}dx
 \label{postad:defup2}
\end{equation}

Let us consider a (reasonable) approximation that the distribution function for particles undergoing acceleration is $f(p,x)=f\rs{o}(p){\cal H}(x\rs{p}-x)$ where $f\rs{o}(p)$ is the distribution at the shock and ${\cal H}(x)$ is the Heaviside step-function, which means that particles with the momentum $p$ which continue acceleration are within their diffusion length $x\rs{p}$ from the shock. 
Then $u\rs{p1}=u_1(x\rs{p1})$ and $u\rs{p2}=u_2(x\rs{p2})$. 

The `local' (in the momentum space) index of the particle distribution function is given by 
\begin{equation}
 s\equiv-\frac{d \ln f\rs{o}}{d \ln p}=\frac{3u\rs{p1}}{u\rs{p1}-u\rs{p2}}+\frac{d\ln(u\rs{p1}-u\rs{p2})}{d\ln p}.
 \label{postad:spinddef}
\end{equation}

\section{Toy model A. Case of a constant gradient of the flow velocity}
\label{postad:sect-dudx}

Let us examine properties of the particle distribution in a toy model that results in a simple analytic solution. Namely, a shock propagating with a constant speed and a downstream flow with a constant gradient $du/dx$. The CR back-reaction is negligible, and therefore the flow velocity upstream $u_1$ is spatially constant and $u\rs{p1}=u_1$. 
The velocity $u\rs{p2}=u(x\rs{p})$. The flow velocity at a point $x$ downstream is $u(x)=u_2+[du/dx]x$; $u_1>0$ and $u_2=u_1/4>0$ are immediate pre- and post-shock values. 
By substituting Eq.~(\ref{postad:def-xp-appr0}) with this $u(x)$ and considering a spatially constant diffusion coefficient, we derive a quadratic equation for $x\rs{p}$ with a real positive solution 
\begin{equation}
 \frac{x\rs{p}}{R}=\frac{1}{\mu}\left[\sqrt{1+2\mu\frac{\tilde x\rs{p}}{R}}-1\right], 
 \qquad \mu\equiv \frac{[du/dx]}{u_2/R}>-\frac{R}{2\tilde x\rs{p}}
\end{equation} 
where $\tilde x\rs{p}=D/u_2$.  
As an example, the parameter $\mu\approx 1$ for Sedov shock \citep{1950RSPSA.201..159T}. The ratio $\tilde x\rs{p}/R\ll 1$ and we may use the first-order approximation $x\rs{p}\simeq \tilde x\rs{p}\propto p^{-1/\kappa}$.

The particle distribution at the shock for this problem can be obtained in analytic form from the general solution (\ref{postad:f0sol}):
\begin{equation}
 f\rs{o}(p)=\frac{\xi n_1}{4\pi p_0^3}\frac{3u_1}{u_1-u_2}\left[1-\frac{[du/dx]x\rs{p}}{u_1-u_2}\right]^{\,\delta}
 \left(\frac{p}{p_0}\right)^{-3u_1/(u_1-u_2)}
 \label{postad:f0-dudx}
\end{equation}
where $\delta={[(3\kappa-1)\,u_1+u_2]/(u_1-u_2)}$. This is essentially the classic spectrum $f\rs{o}\propto p^{-4}$ for momenta $p$ which correspond to the diffusion lengths 
\begin{equation}
 x\rs{p}\ll \frac{(u_1-u_2)}{|du/dx|} \quad\Leftrightarrow\quad 
  \frac{x\rs{p}}{R}\ll\frac{3}{|\mu|}.
\end{equation}
At momenta relevant to $x\rs{p}/R\gtrsim 0.3/{|\mu|}$, the shape of the particle spectrum depends on the sign of the velocity gradient. 
When $du/dx>0$, the local spectral index $s$ increases with momentum $p$, leading to a kind of high-energy cutoff in the spectrum. Conversely, if $du/dx<0$, the spectral index $s$ decreases with $p$, resulting in a harder spectrum. For example, in the case of the Sedov-Taylor shock, where $du/dx>0$,  the spectrum steepens at very high $p$ \citep[figure~3 in][]{2024A&A...688A.108P}.

\section{Toy model B. Maximum momentum}

\subsection{Numerical solutions of the time-dependent kinetic equation}
\label{postad:sect-tacc-1}

In the present section, we consider another toy model to illustrate and further explore the effect of the radial profile of the flow speed downstream on the particle acceleration. 

Specifically, we use RATPaC (Radiation Acceleration Transport Parallel Code) \citep{2012APh....35..300T,2013A&A...552A.102T,2020A&A...634A..59B} to derive a numerical solution of the time-dependent kinetic equation for particle acceleration with a given flow velocity profile. Namely, we consider an unmodified shock with the flow speed in the laboratory frame as 
\begin{equation}
 v(r)=\left\{
 \begin{array}{ll}
 v\rs{s},& \mathrm{for\ } 0\leq r\leq r\rs{B},\ r\rs{A}\leq r\leq R\\
 v\rs{AB},& \mathrm{for\ } r\rs{B}<r<r\rs{A}\\
 0,& \mathrm{for\ } r>R
 \end{array}%
 \right.
 \label{postad:toy-model-v}
\end{equation}
where $v\rs{s}=3V/4$ is the common post-shock value of the flow speed. We consider the following parameters $v\rs{AB}/V=3$, $r\rs{B}/R=0.97$, $r\rs{A}/R=0.99$. 
This converts to the shock frame as $u=V-v$, $x=R-r$, i.e.
\begin{equation}
 u(x)=\left\{
 \begin{array}{ll}
 u_1,& \mathrm{for\ } x< 0\\
 u\rs{AB},&\mathrm{for\ }x\rs{A}<x<x\rs{B}\\
 u_2,& \mathrm{for\ } 0\leq x\leq x\rs{A},\ x\rs{B}\leq x\leq R
 \end{array}%
 \right.
 \label{postad:toy-model-u}
\end{equation}
where $u_1=V$, $u_2=u_1/4$ with the values $u\rs{AB}=-2u_1$, $x\rs{A}/R=0.01$, $x\rs{B}/R=0.03$.

The solution for such a problem is compared to the solution for the uniform velocity profile $v(r)=v\rs{s}$ for $0\leq r\leq R$. 
The shock speed $V=500\un{km/s}$ and the shock compression factor $\sigma=4$ are constant over time, while the radius is set to vary as $R=Vt$. The particles are continuously injected with a constant injection efficiency $\xi$ at a few MeV. 
The diffusion coefficient is Kolmogorov-like 
$D=2.5\cdot10^{9} (w/3) (pc/qB)^{1/3}\un{cm^2/s}$ where $w$ is the particle speed. 
The strength of the magnetic field is $B_1=5\un{\mu G}$ and $B_2=\sigma\rs{B}\,B_1$ with $\sigma\rs{B}=\sqrt{11}$. The diffusion coefficient changes accordingly when a particle crosses the shock. Simulations start at $t_0=20\un{yrs}$. 

\begin{figure}
  \centering 
  \includegraphics[width=\columnwidth]{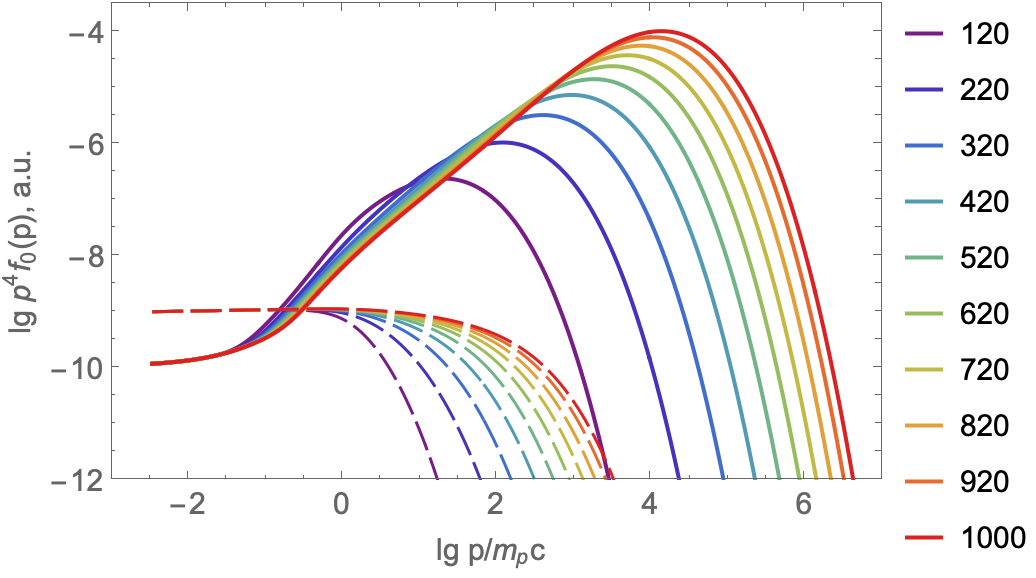}
  \caption{The volume-integrated spectra for particles inside SNR at different times (marked in the plot legend, in years). Solid lines represent the model with parameters $u\rs{AB}=-2u_1$, $x\rs{A}/R=0.01$, $x\rs{B}/R=0.03$. Spectra shown by the dashed lines are calculated for the uniform flow speed. 
  }
  \label{postad:fig-numerical}
\end{figure}

Fig.~\ref{postad:fig-numerical} shows distributions of particles as a function of momentum.  
We emphasize that the model considered exaggerates the effect of the non-uniformity of the downstream flow. In particular, by keeping $x\rs{A}/R$ and $x\rs{B}/R$ constant while the shock expands, we increase the region of negative velocity over time. The particle spectra derived for the toy model are non-physical, with unrealistic energies transferred to CRs. A particular reason is that we set $V(t)=\mathrm{const}$ in the simulations. This is done to isolate the effect of the flow velocity profile on the particle acceleration. In reality, the speed of the post-adiabatic shock decreases efficiently with time. 

There are some features in the particle spectra in the model (\ref{postad:toy-model-v}) noticeable in Fig.~\ref{postad:fig-numerical}.
\begin{itemize}
\item Compared to the case of the uniform flow (represented by dashed lines), one can see that the significantly larger energy is transferred by the shock into the high-energy CRs (solid lines), and this energy grows with time. 
\item The maximum momenta of particles are larger (solid lines) than in the classic case (dashed lines). 
\item After the first spectral break in the spectrum (around $0.1m\rs{p}c$), the spectra are nearly power laws up to some momentum and then, at a higher momentum, the cutoff happens around $p\rs{max}$.
\item We have also calculated a model with a wider region of negative $u_2$, with $x\rs{B}=0.05R$. The maximum momentum is higher if the region with negative $u_2$ is wider. 
\end{itemize}

Let us compare these numerical results with analytical estimates by considering the case with $x\rs{B}=0.03R$. 
Particles scattering back to the shock from distances $x\leq x\rs{p}$ are not affected by the negative $u_2$. Their diffusion length is $x\rs{p}=D_2(p)/u_2$. From this relation, we may estimate the momentum $p\rs{A}$ corresponding to the diffusion length $x\rs{A}=0.01R$. It is $p\rs{A}=0.19m\rs{p}c$. Indeed, the spectra in Fig.~\ref{postad:fig-numerical} (solid lines) deviate from a simple power law with $s=4$ around these momenta. 

\begin{figure}
  \centering 
  \includegraphics[width=\columnwidth]{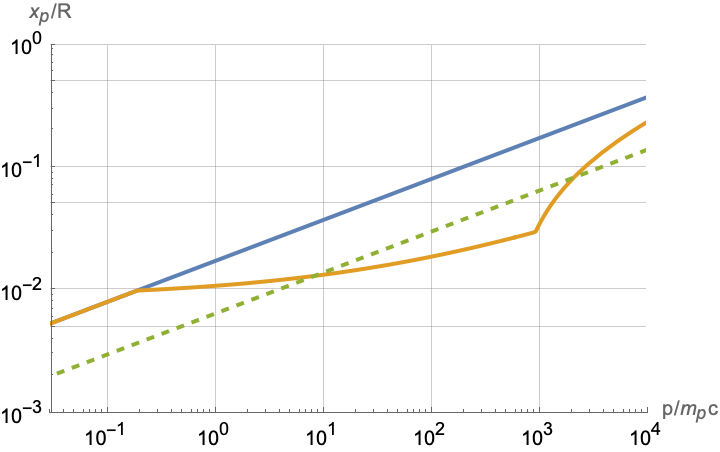}
  \includegraphics[width=\columnwidth]{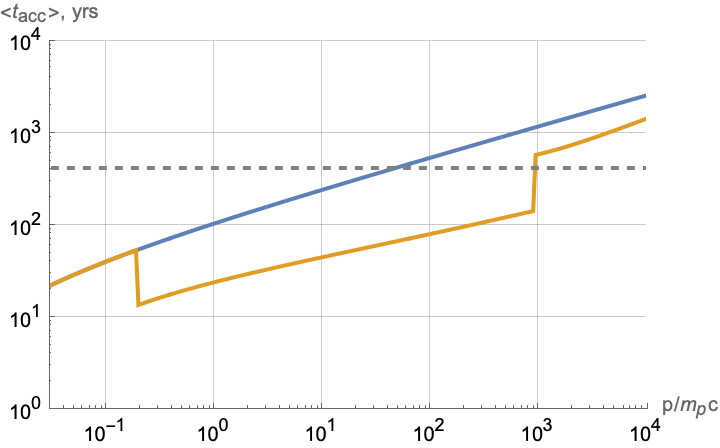}
  \caption{The diffusion length $x\rs{p2}(p)$ (top) and mean acceleration time $\langle t\rs{acc}(p)\rangle$ (bottom) calculated in the analytical approach (i.e., it is time-independent). 
  The model with $x\rs{A}=0.01R$, $x\rs{B}=0.03R$, $u\rs{AB}=-2u_1$ is shown with the orange line;  it is calculated with Eqs.~(\ref{postad:def-xp-appr0}) and (\ref{poastad:eqtacc-xp}). The model with uniform $u_2$ is shown with the blue line, as from $x\rs{p}=D_2/u_2$ (then simply $x\rs{p}\propto p^{1/3}$) and (\ref{poastad:eqtacc-uni-2}) for acceleration time. The green dashed line shows $x\rs{p1}(p)$ upstream, and the gray dashed line corresponds to $t=420\un{yrs}$. 
  }
  \label{postad:fig-numerical-xpp-tacc}
\end{figure}

The situation is different for particles with $p\gtrsim p\rs{A}$ because their $x\rs{p}>x\rs{A}$. The diffusion length calculated with Eq.~(\ref{postad:def-xp-appr0}) for $t=420\un{yrs}$ is shown in Fig.~\ref{postad:fig-numerical-xpp-tacc} (top plot). The momentum $p\rs{B}$ corresponding to $x\rs{B}=0.03R$ is $p\rs{B}=915m\rs{p}c$, as from the analytical formula (\ref{postad:def-xp-appr0}). From Fig.~\ref{postad:fig-numerical}, we see that the numerical spectrum for $t=420\un{yrs}$ indeed deviates from the hard power law (i.e., with a spectral index $s<4$) around this momentum. 

Particles with momenta $p\rs{A}<p<p\rs{B}$ are scattered back to the shock from smaller distances $x\rs{p}$ from the shock (orange line) compared to the case of the uniform $u_2$ (blue line). For example, particles with momentum $p\rs{B}$ return to the shock from a distance $0.17R$ in the model with the uniform flow but from $0.03R$ in the model with the flow as in Eq.~(\ref{postad:toy-model-u}). Clearly, the time needed for one acceleration cycle is shorter in the latter case, and particles can be accelerated to higher $p\rs{max}$ during the same period of time. We consider this further in Sect.~\ref{postad:sect-tacc-2}.

\begin{figure}
  \centering 
  \includegraphics[width=\columnwidth]{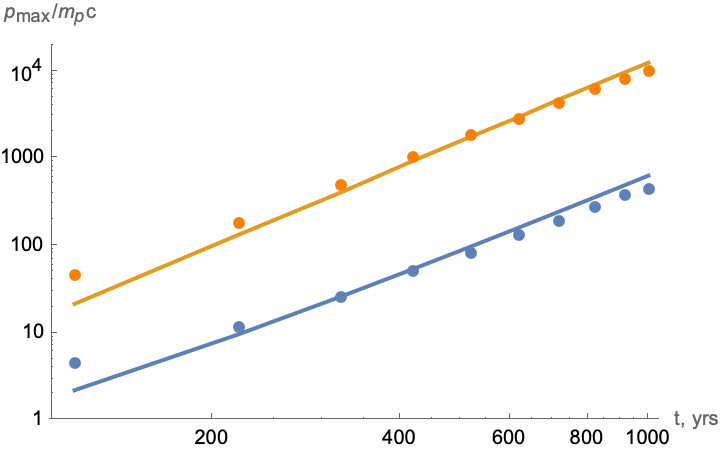}
  \caption{Maximum momenta $p\rs{max}$ for two numerical solutions (dots) and from analytical expressions (solid lines). The blue line is from Eq.~(\ref{poastad:eqtacc-uni-2}) and dots are from the numerical solution with uniform $u_2$ (dashed lines in Fig.~\ref{postad:fig-numerical}). The orange line is calculated with Eq.~(\ref{poastad:eqtacc-xp}) with $x\rs{p1}$ and $x\rs{p2}$ from (\ref{postad:def-xp-appr0}) and dots are from the numerical solution with $x\rs{B}=0.03R$, $u\rs{AB}=-2u_1$ (solid lines on Fig.~\ref{postad:fig-numerical}). 
  The value of $p\rs{max}$ is measured at the level where the distribution function drops below the power-law extrapolation in $e$ times for the orange dots and in $10$ times for the blue dots. 
  }
  \label{postad:fig-numerical-pmax}
\end{figure}

Under the approximation $f(x,p)=f\rs{o}(p){\cal H}(x\rs{p}-x)$, the integral (\ref{postad:f0sol}) for the analytical solution may be separated into three integrals, in the domains $(p\rs{o},p\rs{A})$, $(p\rs{A},p\rs{B})$, $(p\rs{B},p\rs{max})$. The flow velocity is constant in each of these domains. Therefore, the distribution function $f\rs{o}(p)$ should be given by a sequence of power laws with spectral indices $s=4$ until $p\rs{A}$, $s=1$ until $p\rs{B}$, and $s=4$ until $p\rs{max}$. The numerical spectra in Fig.~\ref{postad:fig-numerical} follow this sequence with the addition of the high-energy cutoff, which smooths the last power-law part of the analytical spectrum. Some additional details on the shape of the CR spectrum in a numerical solution are given in Appendix~\ref{postad:appslope}.

In summary, if there is a region $x>x\rs{A}$ with a ratio $u_2/u_1$ different from the value relevant for compression at a strong adiabatic shock $0.25$, then the spectrum of accelerated particles with $p\gtrsim p\rs{A}$ has a spectral index $s=3/(1-u_2/u_1)$ that is different from the classic $s=4$. 
In particular, if $u_2/u_1>0.3$, the index of the distribution function $N(p)\equiv p^2f\rs{o}(p)\propto p^{-\Gamma}$ is $\Gamma\gtrsim 2.3$ as needed to fit gamma-ray spectra in some middle-aged SNRs interacting with dense molecular clouds, like IC443 or W44 \citep[e.g.][]{2013Sci...339..807A}. Indeed, the situation $u_2/u_1>0.3$ occurs if the shock suddenly interacts with a dense material. Then, the shock speed rapidly decreases while the internal material still has a speed corresponding to the pre-interaction times. 

\subsection{Mean acceleration time}
\label{postad:sect-tacc-2}

How can we estimate the acceleration time and the maximum momentum if the flow velocity is not spatially constant?

It is well known that the average acceleration time for the case of spatially constant diffusion coefficient $D$ and flow velocities $u_1$, $u_2$ is \citep{1983RPPh...46..973D}
\begin{equation}
 \langle t\rs{acc}(p) \rangle= \frac{3}{u_1-u_2}\int_{p\rs{o}}^{p} 
 \left(\frac{D_1(p')}{u_1} +\frac{D_2(p')}{u_2}\right)
 \frac{dp'}{p'}	.
\label{poastad:eqtacc-uni} 
\end{equation}
With $D_1$, $D_2$, $u_1$, $u_2$ spatially constant in their domains and the diffusion coefficient $D=\eta cr_0\,(r\rs{g}/r_0)^{1/\kappa}/3$ used in Sect~\ref{postad:sect-tacc-1}, Eq.~(\ref{poastad:eqtacc-uni}) transforms to 
\begin{equation}
 \langle t\rs{acc}(p) \rangle= \frac{3\beta\rs{a}D_1(p)}{V^2}
 , \quad \beta\rs{a}=\frac{\kappa\,\sigma\,\left(1+\sigma/\sigma\rs{B}^{1/\kappa}\right)}{\sigma-1}
\label{poastad:eqtacc-uni-2} 
\end{equation}
for $p\gg p\rs{o}$, where $\beta\rs{a}$ accounts for the fractions of time particles spend upstream and downstream. It is $\beta\rs{a}=14.7$ for $\sigma=4$, $\sigma\rs{B}=\sqrt{11}$ and Kolmogorov diffusion. This dependence is shown in Fig.~\ref{postad:fig-numerical-xpp-tacc} with the blue line. The maximum momentum which may be reached at $t=420\rs{yrs}$ is $p\rs{max}=55 m\rs{p}c$. This corresponds to the maximum momentum in the spectrum from the bottom right plot in Fig.~\ref{postad:fig-numerical}. 

However, the maximum momentum in the spectrum for $t=420\un{yrs}$ is considerably higher in the region where the negative $u_2$ is accounted for (Fig.~\ref{postad:fig-numerical} bottom left plot). Therefore, the formula (\ref{poastad:eqtacc-uni}) should be modified. 

The two ratios under the integral in Eq.~(\ref{poastad:eqtacc-uni}) are actually estimates for $x\rs{p}(p)$. Therefore, we suggest an ansatz for the acceleration time that incorporates the spatial variations of the flow speed and magnetic field: 
\begin{equation}
 \langle t\rs{acc}(p) \rangle\simeq\frac{3}{u_1(x\rs{p1})-u_2(x\rs{p2})}\int_{p\rs{o}}^{p}  \left(x\rs{p1}(p') +x\rs{p2}(p')\right)
 \frac{dp'}{p'}
 \label{poastad:eqtacc-xp}  
\end{equation}
with $x\rs{p}(p)$ given by the equation (\ref{postad:def-xp-appr0}). 
The orange line on the bottom plot in Fig.~\ref{postad:fig-numerical-xpp-tacc} shows the acceleration time calculated with the expression (\ref{poastad:eqtacc-xp}). The acceleration time is shorter for particles with momenta $p>p\rs{A}$ if there is a region with an inverted flow velocity downstream.  

Since acceleration is faster, the maximum momentum is limited by the available time
$\langle t\rs{acc}(p\rs{max})\rangle=t$ may be as high as $915m\rs{H}c$ in $420\un{yrs}$ at the shock with $v\rs{AB}=3V$, while it could reach just $47m\rs{H}c$ at the shock with uniform $u_2$ (Fig.~\ref{postad:fig-numerical-xpp-tacc}).

To demonstrate the reliability of the expression (\ref{poastad:eqtacc-xp}), we compare the maximum momenta it gives with those derived from the numerical solution in Fig.~\ref{postad:fig-numerical-pmax} (orange line and dots). Impressively, we can see an almost perfect correspondence. In parallel, the blue line is calculated with Eq.~(\ref{poastad:eqtacc-uni-2}) and it corresponds to the numerically derived $p\rs{max}$ in the model with uniform $u_2$. 
The numerical solution clearly demonstrates the effect shown by analytical formulae, namely, the higher $p\rs{max}$ if there is a region with a negative flow velocity downstream. 

Finally, the orange line in Fig.~\ref{postad:fig-numerical-pmax} perfectly matches $p\rs{max}\propto t^{\,3}$, as expected. Indeed, the dependence $p\rs{max}\propto t^{\,\kappa}$ (with $\kappa=3$ for the Kolmogorov diffusion coefficient) follows from a simple relation $t\rs{acc}\propto D(p)\propto p^{1/\kappa}$.

\begin{figure}
  \centering 
  \includegraphics[trim=15 80 15 15,clip,width=\columnwidth]{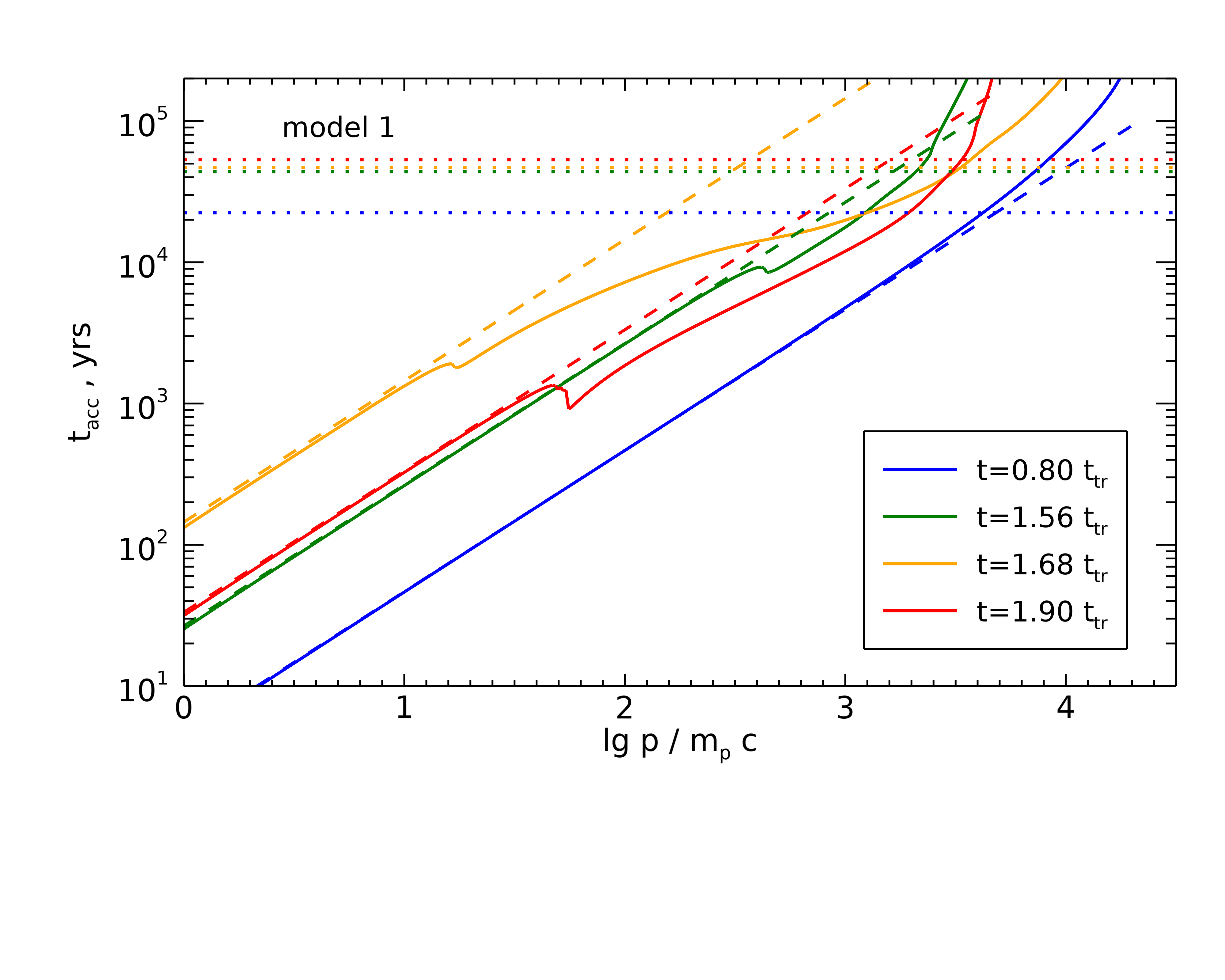}
  \caption{The mean acceleration time $\langle t\rs{acc}(p)\rangle$ for our numerical  model 1 and Bohm diffusion coefficient.
  Solid lines show calculations with Eqs.~(\ref{poastad:eqtacc-xp}) and (\ref{postad:def-xp-appr0}) for $x\rs{p}$. 
  Lines of different colors are calculated with the spatial distributions of MHD parameters relevant to the time moments $t$ shown in the legend.  
  The dashed lines of the same color correspond to the same time moments and uniform flow, Eq.~(\ref{poastad:eqtacc-uni-2}) with $\beta\rs{a}=6.67$.
  Dotted horizontal lines are added to guide the eye; they correspond to the time moments shown in the legend.
  }
  \label{postad:fig-tacc-hdmodels}
\end{figure}
\begin{figure}
  \centering 
  \includegraphics[trim=15 80 15 15,clip,width=\columnwidth]{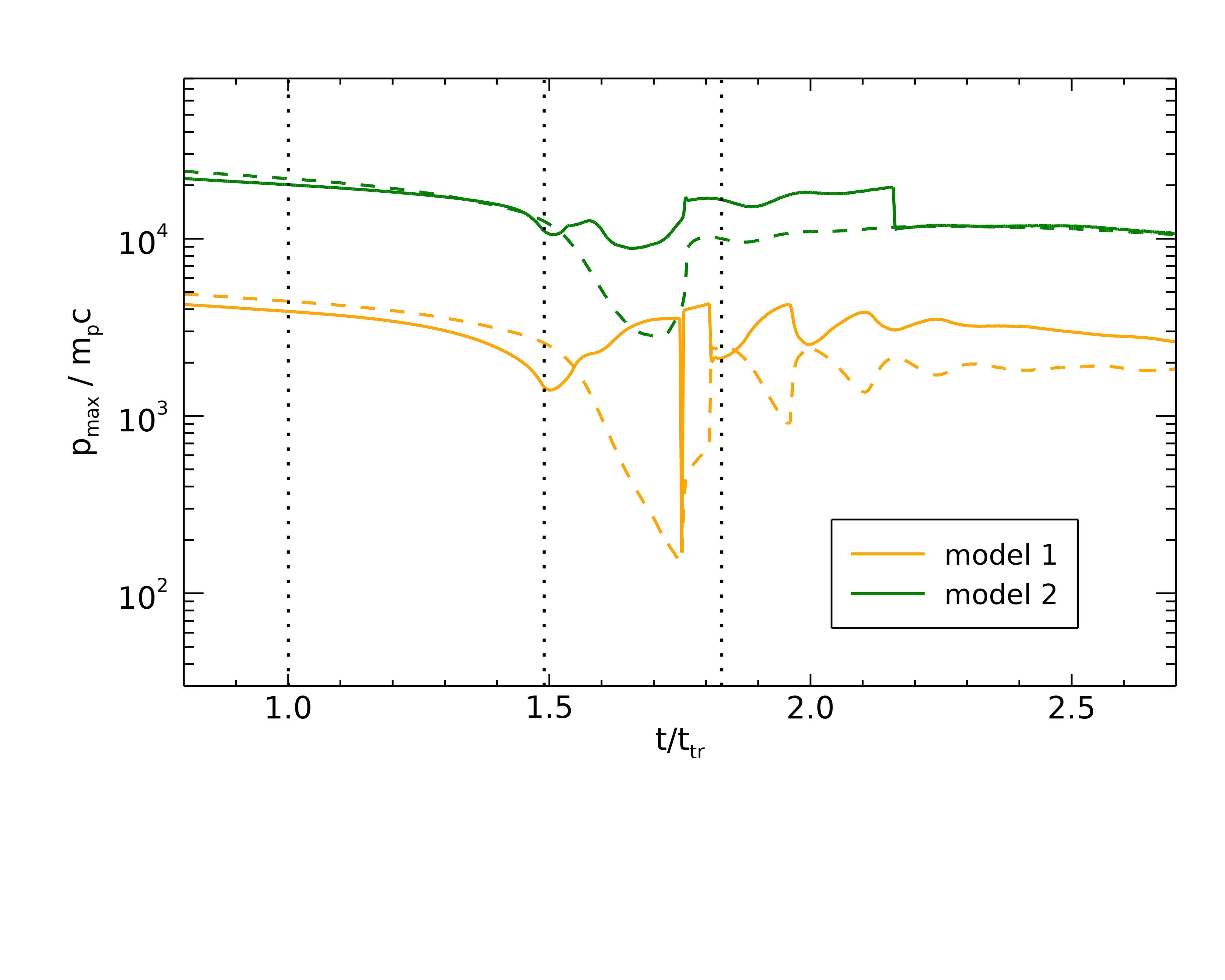}
  \caption{The time-limited maximum momentum for our MHD models and Bohm diffusion coefficient. As in Fig.~\ref{postad:fig-tacc-hdmodels}, the solid lines represent calculations with the spatial structure of the flow, while the dashed lines are for the uniform flow. 
  Three vertical lines mark the times $t\rs{tr}$ (left), $1.49t\rs{tr}$ (when the region with negative $u_2$ appears) and $t\rs{sf}$ (right).
  }
  \label{postad:fig-pmax-hdmodels}
\end{figure}

\begin{figure}
  \centering 
  \includegraphics[trim=20 80 15 15,clip,width=\columnwidth]{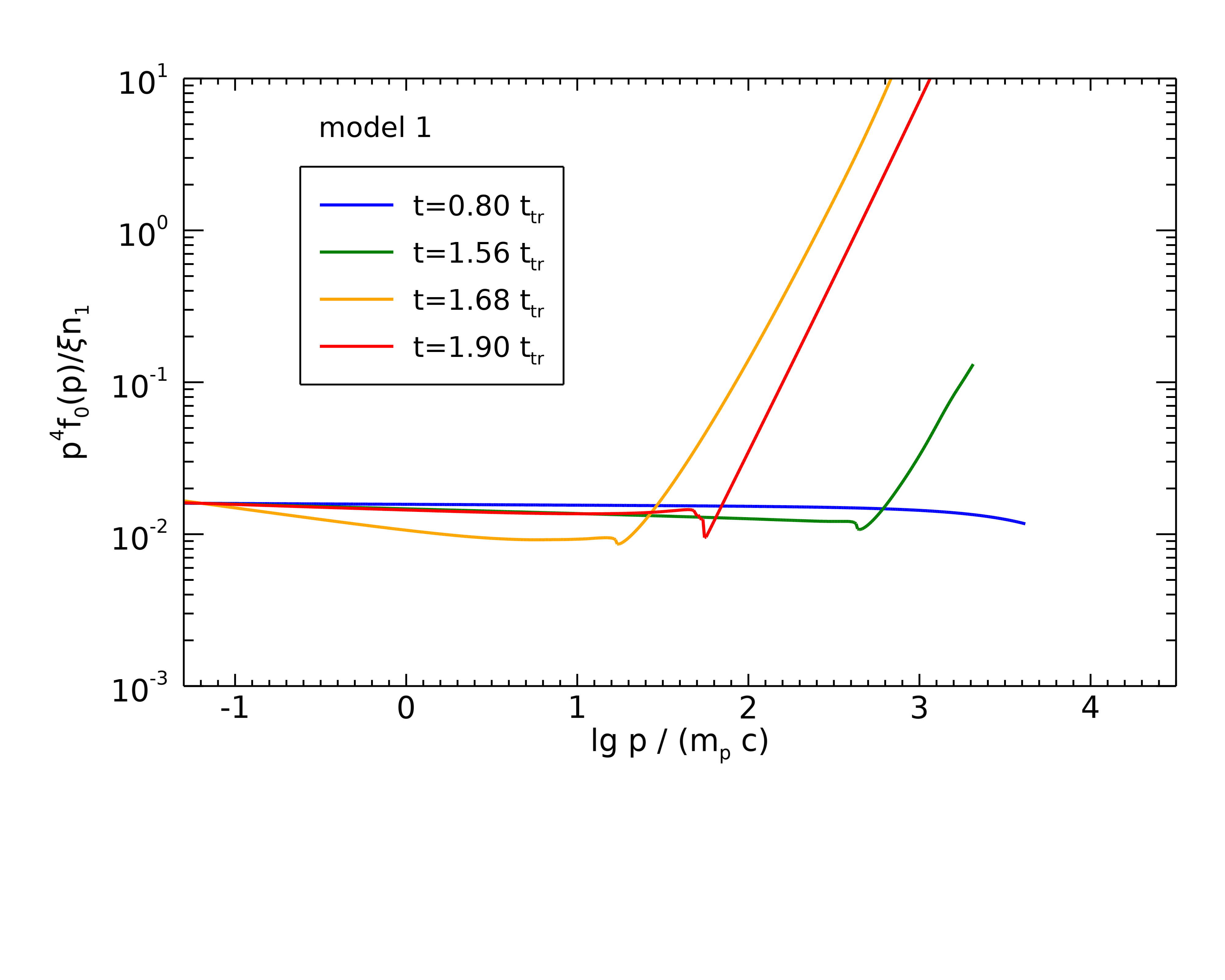}
  \includegraphics[trim=20 80 15 15,clip,width=\columnwidth]{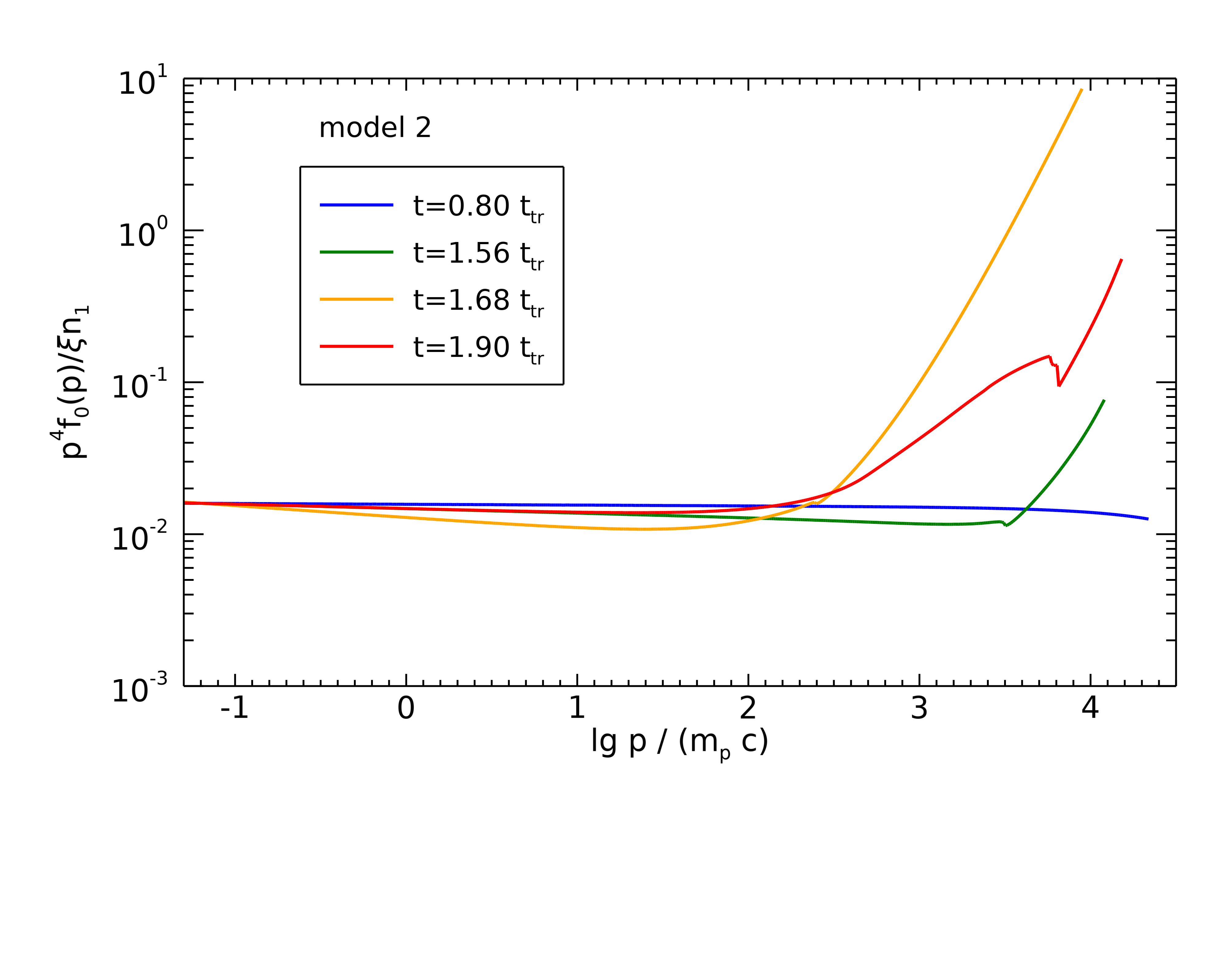}
  \caption{Spectra of accelerated particles for the model 1 and model 2 with Bohm diffusion obtained with equations from Sect.~\ref{postad:sect-f0}. Color lines correspond to different time moments.
  }
  \label{postad:fig-f0s-5}
\end{figure}
\begin{figure}
  \centering 
  \includegraphics[trim=25 80 10 15,clip,width=\columnwidth]{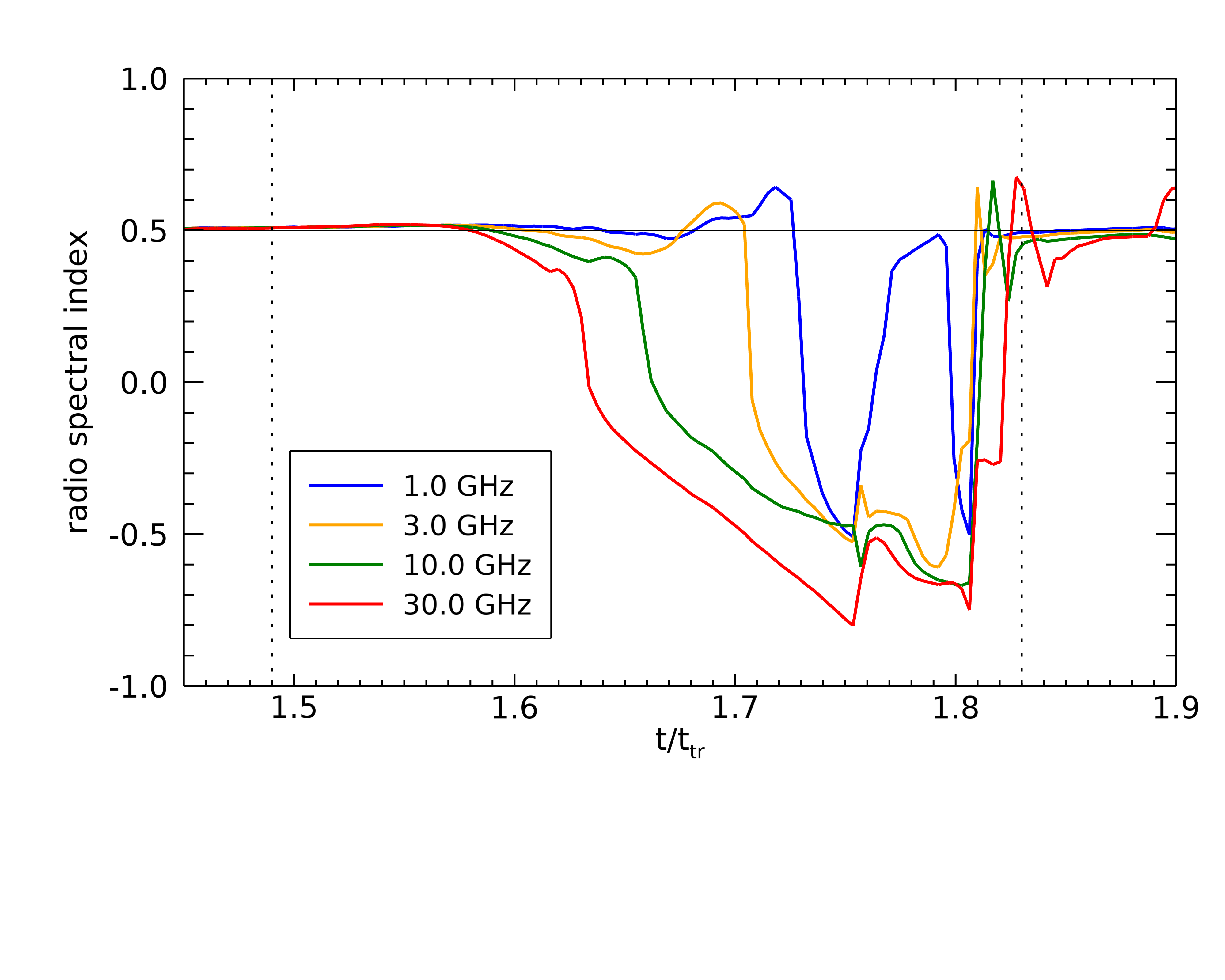}
  \includegraphics[trim=25 80 10 15,clip,width=\columnwidth]{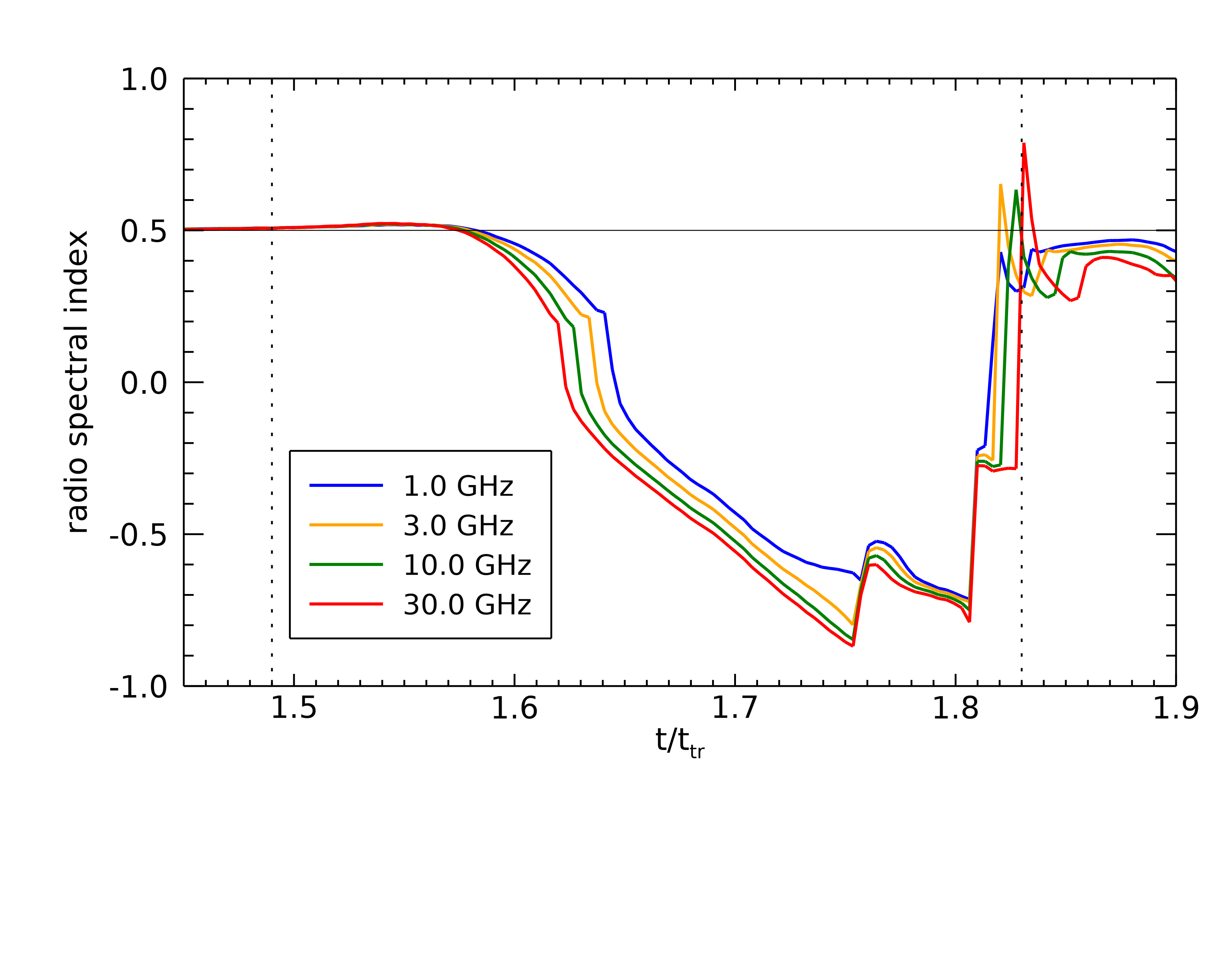}  
  \caption{The evolution of the radio index for model 1 at a few frequencies shown in the legend, in GHz.
  Bohm (\textit{top}) and Kolmogorov (\textit{bottom}) diffusion are assumed.
  We set $D\rs{Kolm}(p\rs{*})=D\rs{Bohm}(p\rs{*})$ at $p\rs{*}=55m\rs{p}c$ for this plot.
  Vertical lines mark the time $t/t\rs{tr}=1.49$ (then the negative $u_2$ first appears) and $t\rs{sf}$.
  }
  \label{postad:fig-radio-s}
\end{figure}

\section{Evolution of spectra in the MHD models}

In the present section, we return to our MHD models, namely model 1 and model 2.

Fig.~\ref{postad:fig-tacc-hdmodels} shows the mean acceleration time $\langle t\rs{acc}(p)\rangle$ calculated for our model 1 in the two approaches. The solid lines correspond to the detailed calculations with Eqs.~(\ref{poastad:eqtacc-xp}) and (\ref{postad:def-xp-appr0}) which account for the spatial profiles of MHD parameters. The dashed lines represent calculations with Eq.~(\ref{poastad:eqtacc-uni-2}) assuming the spatially uniform flow. 
A given line corresponds to a system configuration (the shock speed, profile of the flow velocity, etc.) at the time indicated in the legend. The line shows the time a particle needs to reach a given momentum. For SNR in the Sedov phase (blue lines), the differences due to assumed uniform and non-uniform fluid velocity (cf. the solid and dashed lines) become apparent only for large momenta, $p/m\rs{p}c\gtrsim 10^4$. When SNR is in the post-adiabatic phase (green lines), the differences between the two approaches appear for lower momenta, $p/m\rs{p}c\gtrsim 10^{2.6}$. These differences affect as low momenta as $p/m\rs{p}c\gtrsim 30$ for later evolutionary epochs (the yellow and the red lines).
There are two effects apparent from this plot. 
On the one hand, by comparing the lines with each other, we see that as the system enters the radiative phase, the acceleration time for most momenta first increases (blue, green, and yellow lines) and then decreases (red lines). This behavior is primarily caused by the temporal deceleration and subsequent acceleration of the shock, since the same trend is observed in both approaches, for the sequences of solid and the sequence of dashed lines.
On the other hand, by comparing a solid with a corresponding dashed line, the flow non-uniformity downstream of the post-adiabatic shock shortens the acceleration time compared to the case of a uniform flow. Indeed, let us consider, for example, the yellow solid and dashed lines and compare the points of their intersection with the yellow dotted horizontal line. This leads to a conclusion that the system with the flow non-uniformity (solid line) is able to accelerate particles to the momentum $p/m\rs{p}c\simeq 10^{3.5}$ while the system with a uniformity to $p/m\rs{p}c\simeq 10^{2.5}$ only.

Fig.~\ref{postad:fig-pmax-hdmodels} shows the maximum momenta derived as a solution of equation $t=\langle t\rs{acc}(p\rs{max})\rangle$, on the flow structures relevant for a given time $t$. Although both approaches (with uniform and non-uniform flow) typically result in similar values of $p\rs{max}$ before $\approx 1.5t\rs{tr}$, the differences may reach an order of magnitude during the post-adiabatic phase, due to impact from the region with negative $u_2$. 

The evolution of the particle distribution function $f\rs{o}(p)$ for the two models is shown in Fig.~\ref{postad:fig-f0s-5} and in animations in Appendix~\ref{postad:app-movie-f0}. 
Similarly to a non-linear acceleration regime, the spectrum becomes concave, though we neglect the back-reaction of CRs on the shock, and $u_1$ is uniform. 
The spectra are so hard because of the quite strong gradients of $u_2$, high differences between $u_1$ and $u_2$, and the presence of the region with $u_2$ directed toward the shock (Fig.~\ref{postad:fig-MHDprofiles}), as considered in detail in Sections~\ref{postad:sect-CRacc} and \ref{postad:sect-dudx}. 

From spectra $f\rs{o}(p)$ at different times, we may calculate the evolution of the spectral index of the radio emission generated by electrons accelerated at the forward shock.
Fig.~\ref{postad:fig-radio-s} shows the evolution of the radio spectral index for the post-adiabatic shock for model 1. It is calculated as described in Appendix~\ref{postad:radio-idx-calc}, considering the emission from a thin shell with varying $B_2(x)$ and $n_2(x)$. 
From Fig.~\ref{postad:fig-radio-s} we can see that the radio spectral index decreases starting from time about $1.6t\rs{tr}$. This is in perfect agreement with expectations and observational hints presented in Sect.~\ref{postad:obs-evid}. 

In the case of Bohm diffusion (top plot), the higher frequencies respond earlier to radiative cooling because of the higher energies of electrons emitting at these frequencies, so they can diffuse to larger distances from the shock (cf. Fig.~\ref{postad:fig-MHDprofiles}). In a similar fashion, the `sensitivity' of the radio index to the structure of the flow is higher for the Kolmogorov diffusion coefficient (bottom plot) because the electrons in this case may probe the deeper regions downstream of the shock. 

The radio spectral index at these frequencies is close to $0.5$ all the time in model 2, either with the Bohm or Kolmogorov diffusion coefficient. The reason is that in this model, particles become sensitive to the flow structure only at higher momenta (Figs.~\ref{postad:fig-Fxx-appr0} and \ref{postad:fig-f0s-5}). By increasing the Kolmogorov diffusion coefficient, the behavior of the radio spectral index in model 2 can be made similar to that in model 1.

\section{Discussion and Conclusions}

The main findings of the present paper can be summarized as follows.
\begin{itemize}
\item MHD simulations of SNRs demonstrate essential changes in the flow structure behind the shock on the post-adiabatic stage. 
\item The changes also affect the profiles of the flow velocity on the length scales of the CR acceleration.
\item As a result, the acceleration efficiency increases. This happens due to three reasons: a higher $u_1/u_2$, an increased $\Delta p$ per acceleration cycle, and a possible existence of a region with $u_2<0$.
\item Together, these factors make the particle spectrum harder. 
\item It could also result in a decrease of the radio spectral index as the shock progresses into the post-adiabatic regime. 
\item MF becomes dynamically important in the post-adiabatic shocks; it affects the profiles of the flow velocity and, therefore, could influence the particle spectrum.
\end{itemize}

We also report the observed data to support these conclusions: the correlation of the radio index in Kes~73 with the interstellar density and the decrease in the radio index with age in interacting SNRs.

However, the application of our theoretical considerations to 1D numerical simulations of the post-adiabatic SNRs seems to exaggerate the overall effect. 
There are a few limitations that one has to take into account. 

In particular, fluctuations in the shock speed $V$ and the parameter $m$ (Fig.~\ref{postad:fig-m}) seem to appear because the numerical MHD simulations are 1D. In 3D, the flow has more degrees of freedom, and we may expect, in particular, that the first minimum in $V$ and $m$ around time $1.7t\rs{tr}$ could not be so deep. One might also expect that the spatial profiles of the flow velocity in more realistic 3D simulations could have softer gradients and contrasts. However, simulations of post-adiabatic flows, in order to be coupled to the CR acceleration, require a resolution down to $\sim 10^{-5}R$ around the shock. Such simulations are very demanding in three dimensions, considering the spatial and temporal ranges that must be covered. 

Another point is that we used the steady-state solution for particle acceleration on a stationary MHD background. This implicitly assumes that the acceleration time is shorter than the hydrodynamic time-scale for spatial changes in the flow. However, once the radiative losses of plasma appear, the structure of the flow changes rapidly (animations in Appendix~\ref{postad:app-movie-MHD}). 
Therefore, though the stationary solution could be reasonable for the radio-emitting electrons, the spectra of CRs at the highest energies (emitting X- and gamma-rays) could be altered in a time-dependent treatment which considers non-stationary acceleration of particles on the non-stationary profiles of the flow velocity. The numerical treatment of such a problem, together with high-resolution 3D MHD simulations, is almost unfeasible at the present time. 

Post-adiabatic shocks, despite their rather low shock speed, could be efficient CR accelerators. Our simulations show that a large amount of energy is transferred into the high-energy CRs (Fig.~\ref{postad:fig-f0s-5}). 
Numerically, the energy in CRs (as from an instant $f\rs{o}(p)$) reaches $50\%$ of the shock kinetic energy about time $1.6t\rs{tr}$ and rapidly grows from that time on. 
If the energy in CRs becomes this high, then their back-reaction on the flow upstream should also be taken into account. 

Furthermore, such high efficiency in transferring kinetic energy into the CRs creates a new channel of energy losses of plasma, in addition to the radiative losses. This should affect both the MHD structure of the flow and the shock speed. 
Thus, the shock decelerates in the post-adiabatic regime not only due to the expansion and losses of thermal radiation but also due to the efficient acceleration of CRs. 
The influence of CRs on the post-adiabatic plasma has recently been considered by \citet{2025ApJ...980..167D}, by introducing a free parameter which sets the ratio of CR pressure to the thermal pressure. Indeed, the authors have shown that CRs can strongly affect the plasma dynamics at the post-adiabatic phase.  

The radiative energy losses, as well as efficient transfer of the shock energy into CRs, should lead to a rapid, successive drop in acceleration efficiency and the disappearance of a post-adiabatic SNR as a non-thermal source.  
In fact, there is evidence of a fade in radio emission in SNRs shortly after the end of the Sedov stage, which comes from an analysis of the $\Sigma-D$ relation \citep{2010A&A...509A..34B}. 

When this paper was under review, a publication has appeared at {\sl arXiv} exploring the CR spectra at shocks with radiative losses \citep{2025A&A...704A.213C}. The author has also demonstrated that the shock during the early stage of the radiative phase could be a significantly efficient CR accelerator.

A more refined future treatment of the problem should involve multi-dimensional MHD simulations coupled with particle acceleration.

\begin{acknowledgements} We thank the referee, Pat Slane, for constructive comments.
We acknowledge Yang Chen for providing us with the ${}^{12}$CO ($J=1\rightarrow 0$) maps. We acknowledge the support from INAF 2023 RS4 Theory grant as well as INAF 2023 RS4 mini-grant. OP and TK thank the Armed Forces of Ukraine for providing security to perform this work. This research used an HPC cluster at DESY, Zeuthen, Germany, and the HPC system MEUSA at INAF-Osservatorio Astronomico di Palermo, Italy.
\end{acknowledgements}

\bibliographystyle{aa}
\bibliography{postadacc} 

\begin{appendix}  

\section{Animations}
\label{postad:app-movie}

\subsection{Evolution of the flow downstream of the post-adiabatic shock}
\label{postad:app-movie-MHD}

The animations demonstrate the time development of the 
profiles of density (upper right), flow speed $v(r)$ in the laboratory frame (lower left), and magnetic field (lower right) for two models of SNRs described in Sect.~\ref{postad:MHD-models}. Namely, \textit{Animation 1} for model 1 with parallel shock and $B\rs{o}=5\un{\mu G}$ and \textit{Animation 2} for model 2 with perpendicular shock and $B\rs{o}=10\un{\mu G}$.
Note that the horizontal axis in the plot for the flow speed spans only $15\%$ of the shock radius, to emphasize changes around the region where the shock particle acceleration happens. The upper left plot shows the expansion parameter and the moving pointer, which indicates the time moment for which the hydrodynamic profiles have been shown.

\subsection{Evolution of the spectrum of particles accelerated at the post-adiabatic shock}
\label{postad:app-movie-f0}

\textit{Animation 3} for model 1 and \textit{Animation 4} for model 2 show the evolution of the flow velocity in the reference frame of the shock $u_2(x)$, the spectrum of accelerated particles $f\rs{o}(p)$ and the spectral index $s(p)=-d\ln f\rs{o}/d\ln p$. 

\section{Some considerations about Kes~73}
\label{postad:appkes73}

By assuming the uniform expansion and considering the Sedov shock in a uniform ambient medium, \citet{2017ApJ...846...13B} have derived 
the average shock speed in Kes~73 $V\simeq 1400\un{km/s}$, 
the average radius $R=6\un{pc}$, 
the average ambient number density $n\rs{o}\simeq 2\un{cm^{-3}}$, the explosion energy $1.9\un{foe}$ and the age $t=1800\un{yrs}$. As we see (Fig.~\ref{postad:fig-radio-s}), the radio spectral index is expected to begin decreasing due to radiative losses before $1.6t\rs{tr}$. 
What should be the ambient density $n\rs{mc}$ to allow for a part of the SNR shock to be at the stage corresponding to $1.6t\rs{tr}$ with $t\rs{tr}=2.83\E{4}E\rs{51}^{4/17}n\rs{1}^{-9/17}\un{yrs}$? Writing $t=1.6t\rs{tr}$ and substituting the explosion energy into the equation for the transition time, we derive the density $n\rs{mc}\simeq 580\un{cm^{-3}}$. This value could naturally be in a molecular cloud near Kes~73. 
Indeed, the column density of material SNR interacts with on the West is $N\rs{H_2}\simeq 10^{22}\un{cm^{-2}}$ \citep{2017ApJ...851...37L}. The size of a cloud with such $N\rs{H_2}$ 
and $n\rs{mc}$ is $\simeq5.6\un{pc}$ along the line of sight. From Fig.~\ref{postad:fig-kes73} the thickness of a cloud filament in the plane of the sky is about the SNR radius, that perfectly matches the estimated size of the cloud along the line of sight. 
With such density, the shock speed in the interaction region drops from the average $1400\un{km/s}$ to the local value $1400\,(n\rs{o}/n\rs{mc})^{1/2}\approx 82\un{km/s}$. Then the post-shock plasma temperature is about $T=14V\rs{km/s}^2\simeq 9\E{4}\un{K}$ that corresponds to the maximum of the cooling curve $\Lambda(T)$. And finally, what is the area of the shock surface interacting with such dense material? It is the ratio of the area of the spherical cap to the area of a sphere:
\begin{equation}
\Omega=\frac{2\pi R^2}{4\pi R^2}\left(1-\sqrt{1-\left(\frac{\varrho}{R}\right)^2}\right)\simeq 0.03
\end{equation}
where $\varrho$ is the average radius of the red area on the spectral index map. 
It appears that the interacting region is a quite small fraction of the SNR surface. In addition, this region is projected on the SNR interior. Therefore, it does not significantly affect the estimates of the average values derived by \citet{2017ApJ...846...13B}. 
All these considerations support our idea that the radio spectral index variation across Kes~73 could mark the onset of radiative losses in the interacting part of the SNR. Nevertheless, this idea remains a hypothesis until a dedicated study.  
In particular, such a study would help discriminate it from other potential effects. For example, the shock compression can be sufficiently high in radiative shocks to produce prominent free-free emission in the radio band, which would flatten the spectrum.

\section{Shape of the numerical CR spectrum}
\label{postad:appslope}

The numerical solution for the momenta $p\gtrsim p\rs{A}$, as measured from Fig.~\ref{postad:fig-numerical}, is nearly a power law $p^4f\rs{*}(p)\propto p^{a}$ with $a$ from $1.7$ for $t=120\un{yrs}$ to $1.2$ for $t=1000\un{yrs}$. This corresponds to the spectral index 
$s\rs{*}=4-a$ (for the function $f\rs{*}(p)\propto p^{-s\rs{*}}$) in the range $2.3\div 2.8$. The analytical solution gives $s=3/(1-u\rs{AB}/u_1)=1$ between $p\rs{a}$ and $p\rs{B}$. Thus, we have $s\rs{*}-s\approx 1.3\div 1.8$. Do the numerical and analytical solutions correspond to each other?

The CR spectra from the numerical solution (Fig.~\ref{postad:fig-numerical}) are integrated over the entire volume. One should take into account differences between the analytical (Sect.~\ref{postad:sect-f0}) and the numerical (Sect.~\ref{postad:sect-tacc-1}) treatment of the CR spectra. The latter is a non-stationary solution for expanding spherical shock, and the former, instead, is a steady-state solution for a plane-parallel shock on an immediate MHD background. 

The volume-integrated spectrum represents particles injected at different times. The injection efficiency $\xi$ (a fraction of the density) does not change over time in both approaches. However, in the numerical solution, the number of injected particles increases over time due to the increase in the spherical shock surface $\propto (R(t)/R_0)^2$. The number density of particles injected at some time moment $t\rs{i}$ decreases over time $t$ as $(R(t)/R(t\rs{i}))^{-3}$ because the volume increases. So, the number density of particles accelerated until the time $t$ to some momentum $p$ is smaller than the number density of particles freshly injected at this time $t$. 
The shock surface at the time $t\rs{i}$ when particles were injected was smaller, and the volume has increased since then. As a consequence, the slope of the numerical spectrum is softer: there are more particles with low momenta just because of the expanding spatial domain. 
To compare with the analytical solution, we should apply a correction to the numerical solution by making the early injection artificially larger: $\xi(t\rs{i})\rightarrow\xi\cdot[R(t)/R(t\rs{i})]^5$ where $t>t\rs{i}$. 

The application of this correction is not straightforward. The CR spectrum at a given time $t$ covers a wide range of momenta. Particles that have some momentum $p$ were injected at different times $t\rs{i}$. However, if we assume that all particles with momentum $p$ spend the same time for acceleration to this momentum, then we may write an approximate relation
\begin{equation}
 f\rs{o}(p)\simeq f\rs{*}(p,t)\cdot[R(t)/R(t\rs{i}(p))]^5
\end{equation} 
where $f\rs{o}(p)$ and $f\rs{*}(p)$ are the analytical and numerical distribution functions, 
$R(t)=Vt$,
$t\rs{i}(p)=t-t\rs{acc}(p)$. 
This transforms to a relation for the spectral indices
\begin{equation}
 s(p)\simeq s\rs{*}(p,t)-\Delta s(p,t),\quad
 \Delta s=\frac{5}{3}
 \left[\frac{t}{t\rs{acc}(p)}-1\right]^{-1}
\end{equation} 
where $t\rs{acc}\propto p^{1/3}$ is used, $s$ and $s\rs{*}$ are indices for $f\rs{o}(p)$ and $f\rs{*}(p)$ respectively. 
Note that $t\geq t\rs{acc}$ and, therefore, $\Delta s\geq 0$ as needed.
If $t\rs{acc}\ll t$, then $\Delta s\rightarrow 0$. On the other hand, if $t\rs{acc}\rightarrow t$, then $\Delta s$ could be quite large.
For intermediate ratios $t/t\rs{acc}\sim 2$, the difference $\Delta s\sim 5/3$ is right about the differences $s\rs{*}-s$ we observed in our results.

\section{Calculation of the radio spectral index}
\label{postad:radio-idx-calc}

Synchrotron emissivity is 
\begin{equation}
    P(\nu)=\int\limits_{0}^{\infty}N(p)\epsilon(p,\nu)dp
\end{equation}
where $p$ is the electron momentum, $N(p)=4\pi p^2f\rs{o}(p)$ is the electron momentum distribution, $\nu$ is the synchrotron frequency, $\epsilon$ is the single-electron emissivity
\begin{equation}
    \epsilon= c_1\mu BF(\nu/\nu_c),\qquad \nu\rs{c}=c_2\mu Bp^2,
\end{equation}
where $c_1=\sqrt{3}e^3/(m\rs{e}c^2)$, $c_2=3e/(4\pi m\rs{e}^3c^3)$.
We take the geometrical factor $\mu=1$, i.e., MF is perpendicular to the line of sight. 
We use the `delta-function' approximation for the special function
\begin{equation}
    F(x)=\frac{8\pi}{9\sqrt{3}}\delta\left(x-0.29\right).
\end{equation}
In this approximation  
\begin{equation}
    P(\nu)=c_3 \nu^{1/2} B^{1/2}
           N\left(p\rs{eff}\right),
           \qquad 
           p\rs{eff}=\left(\frac{\nu}{0.29c_2\mu B}\right)^{1/2}
   \label{kepler:peff}        
\end{equation}
where $c_3$ is a constant, $p\rs{eff}$ is the momentum of the electron that emits most of its energy at frequency $\nu$ in the magnetic field $B$. For example, this approximation yields the well-known dependence $P\propto B^{(s+1)/2}\nu^{-(s-1)/2}$ for $N\propto p^{-s}$. 

Most of the emission comes from a shell of the thickness about $0.1R$, where MHD parameters vary in space.  
The flux density is therefore proportional to an integral of emissivity $P$ over this shell:
\begin{equation}
 S(\nu)\propto
 \nu^{1/2}\int\limits_{0}^{0.1R} B_2(x)^{1/2}
           n_2(x)\,f\rs{o}\big(p\rs{eff}(x,\nu)\big)\,p\rs{eff}(x,\nu)^2\,dx.
\end{equation}
where $p\rs{eff}$ depends on $x$ through $B_2(x)$. The spatial profile of advected particles downstream is approximated as $f(x,p)\approx n_2(x)f\rs{o}(p)/n\rs{2}(0)$ \citep[][equation A.6]{2007A&A...470..927O}.

The radio spectral index at the frequency $v_0$ is determined from fluxes at two frequencies
\begin{equation}
 \alpha(\nu_0)=-\frac{\lg\big[S(\nu_1)/S(\nu_2)\big]}{\lg\left(\nu_1/\nu_2\right)}.
\end{equation}
and we choose these frequencies as 
\begin{equation}
 \frac{\nu_2}{\nu_0}=\frac{\nu_0}{\nu_1}=1.05.
\end{equation}

We note that a simpler approach, namely, $\alpha=(s-3)/2$ where the index $s$ of the electron momentum distribution is measured at momentum $p\rs{eff}(0,\nu_0)$, results in a similar temporal variation of the radio spectral index as in a more complex and accurate approach described above.

\section{Erratum}

We noticed misprints in the text of Appendix~C in the paper \citet{2024A&A...688A.108P}. 

In the sentence between equations (C.1) and (C.2), there should be ``$f(x,p)=0$ at $x=-\infty$ and $\partial f/\partial x=0$ at $x=\pm\infty$'' instead of ``$f(x,p)=0$ at $x=\pm\infty$ and $\partial f/\partial x=0$ at $x=-\infty$''.

Expression (C.7) should read as follows: 
\begin{equation}
 \left[D\pd{f}{x}\right]_{+\infty}-u_3f_3\equiv -u\rs{x}f\rs{o}
\end{equation}

\end{appendix}  

\end{document}